# Classification of Machine Learning Frameworks for Data-Driven Thermal Fluid Models

Chih-Wei Chang and Nam T. Dinh

Department of Nuclear Engineering
North Carolina State University, Raleigh NC 27695-7909
cchang11@ncsu.edu, ntdinh@ncsu.edu

## Abstract

Thermal fluid processes are inherently multi-physics and multi-scale, involving mass-momentum-energy transport phenomena at multiple scales. Thermal fluid simulation (TFS) is based on solving conservative equations, for which – except for "first-principles" direct numerical simulation – closure relations (CRs) are required to provide microscopic interactions or so-called sub-grid-scale physics. In practice, TFS is realized through reduced-order modeling, and its CRs as low-fidelity models can be informed by observations and data from relevant and adequately evaluated experiments and high-fidelity simulations. This paper is focused on data-driven TFS models, specifically on their development using machine learning (ML). Five ML frameworks are introduced including physics-separated ML (PSML or Type I ML), physics-evaluated ML (PEML or Type II ML), physics-integrated ML (PIML or Type III ML), physics-recovered (PRML or Type IV ML), and physics-discovered ML (PDML or Type V ML). The frameworks vary in their performance for different applications depending on the level of knowledge of governing physics, source, type, amount and quality of available data for training. Notably, outlined for the first time in this paper, Type III models present stringent requirements on modeling, substantial computing resources for training, and high potential in extracting value from "big data" in thermal fluid research.

The current paper demonstrates and investigates ML frameworks in three examples. First, we utilize the heat diffusion equation with a nonlinear conductivity model formulated by convolutional neural networks (CNNs) and feedforward neural networks (FNNs) to illustrate the applications of Type I, Type II, Type III, and Type V ML. The results indicate a preference for Type II ML under deficient data support. Type III ML can effectively utilize field data, potentially generating more robust predictions than Type I and Type II ML. CNN-based closures exhibit more predictability than FNN-based closures, but CNN-based closures require more training data to obtain accurate predictions. Second, we illustrate how to employ Type I ML and Type II ML frameworks for data-driven turbulence modeling using reference works. Third, we demonstrate Type I ML by building a deep FNN-based slip closure for two-phase flow modeling. The results show that deep FNN-based closures exhibit a bounded error in the prediction domain.

# 1. Introduction

## 1.1. Thermal fluid models

Thermal fluid simulation (TFS) involves multi-physics models such as momentum and energy conservation partial differential equations (PDE). While details vary, thermal fluid models used in engineering practice belong to family of reduced-order models (ROMs) that require sub-grid-scale (SGS) physics models [1, 2] to represent microscopic interactions. SGS models are often referred to as "closure relations" (CRs) or constitutive models, as they serve to close PDE-based models. Traditionally, it takes extensive efforts to gain insights and mechanistic understanding through analysis of observations and data from experiments to develop closure models that effectively represent data in a compact form. The long time required for new model development constrains the application of simulation while dealing with new system conditions or newly designed systems. As an alternative to mechanistic and semi-analytical models, machine learning (ML) methodologies become attractive since they can be used to capture the underlying correlation behind data using nonparametric models, or so-called data-driven models.

### 1.1.1. Traditional framework for developing closure relations

Fig. 1 depicts the traditional framework for developing closure models. Starting from the knowledge base, we can design experiments to obtain relevant data, perform data analysis and research to derive sought-after closure models for thermal fluid simulation. This research may take years to decades. The long time needed for developing new closure relations limits the pace of model applications while dealing with new geometries, new system characteristics, and newly designed systems. The so-obtained closure models are implemented into thermal fluid simulation for applications and assessments. Once tested and evaluated, the obtained models enhance and contribute to the knowledge base.

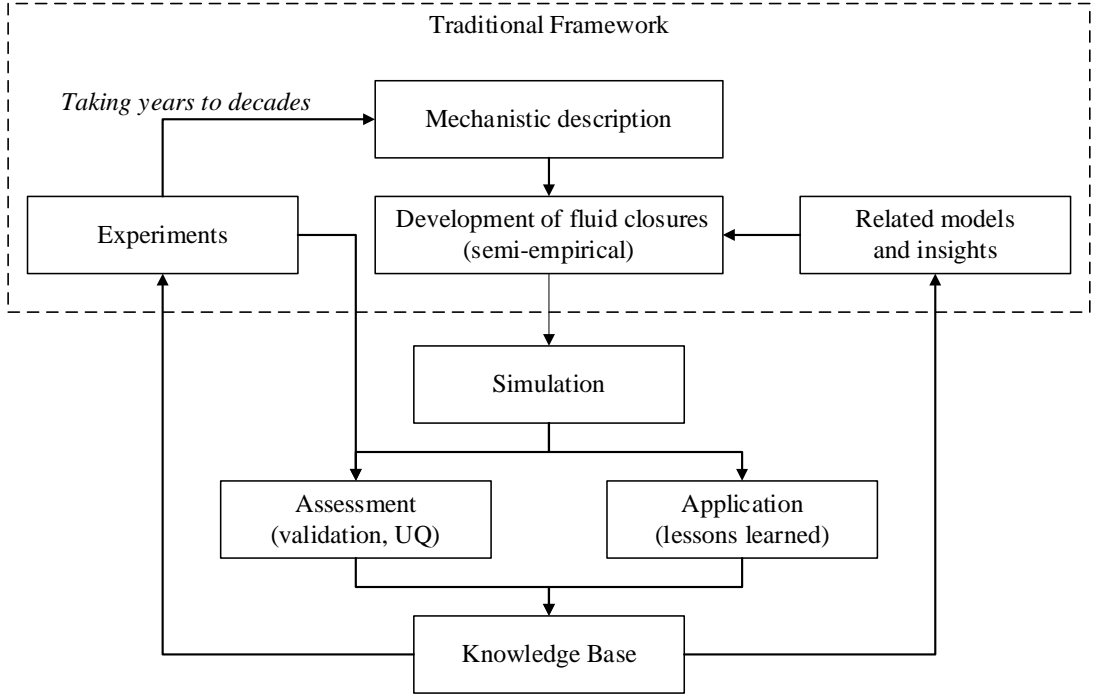

Fig. 1. Traditional framework for developing closure relations.

### 1.1.2.  Data-driven modeling framework for developing closure relations

The research has been pursued in formulating and applying the data-driven modeling (DDM) framework to develop closure models to close conservation equations, particularly leveraging on advances in ML techniques. Nowadays, thermal fluid data accumulate rapidly in a tremendous amount from high-resolution simulations and experiments, primarily due to the affordability of high-performance computing and advances in flow diagnostics, thermal imaging, and other measurement equipment such as high-speed, high-resolution optical and infrared cameras. The value of those high-fidelity data lies with their use (usability) to reduce uncertainty in simulation. The "high fidelity" refers to the data which have been adequately evaluated, and hence trustworthy. Lewis *et al.* [3] investigated a strategy of using high-fidelity data from computational fluid dynamics (CFD) to inform low-fidelity models in system-level thermal hydraulics simulation. They also demonstrated this high-to-low (Hi2Lo) strategy by utilizing a neutron transport equation to inform a neutron diffusion equation. Methodologically, DDM belongs to the Hi2Lo strategy. Its distinctive features relax closure relations from their traditional "mechanistic" models to ML-based models and using ML to extract the value of a substantial amount of data ("big data") for training models.

For the "Big Data" to become useful in DDM, it has to undergo several processing steps [4]. First, results of high-fidelity simulations and experiments need to be collected, categorized, and archived in an easily accessible storage format. Second, the value of data as information needs to be assessed, to establish their relevance to the conditions and models under consideration, so that these data become useful information. Third, data are processed by various methods (including ML) to recognize underlying correlations behind the information. The so-developed intelligence (e.g., in the form of closure relations) is used to enable thermal fluid simulation in applications.

Stemming from the preceding discussion, Fig. 2 depicts the DDM framework that includes the concept of (Data/Method/Platform) "as a Service (aaS)" [5]. The "aaS" notations are employed to denote components (modules) of work-flow in a "divide-and-conquer" strategy that allows us to decompose the framework by different disciplinary fields. We can define requirements, evaluate methods, and review the essential knowledge in each disciplinary field such as machine learning, thermal fluid experiment and simulation, and numerical methods. This concept allows each module to be reused, extended, and improved based on newly observed data as well as newly developed state-of-the-art methods.

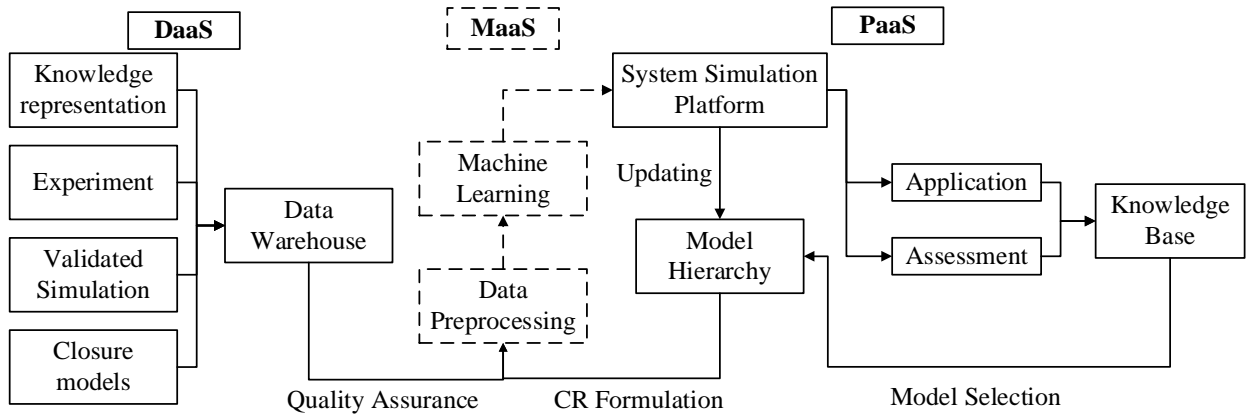

Fig. 2. Overview of the data-driven modeling (DDM) framework.

The DaaS module integrates data from various sources to support closure developments. The MaaS module includes different ML algorithms that can be deployed to infer models from data. The PaaS module contains thermal fluid models that are adaptive to various applications. The detail functions of each module are described:

**Data as a Service (DaaS).** Four types of data need to be stored in the data warehouse including the knowledge representation, experiment, validated simulation, and existing closure models. Experts' knowledge needs to be formalized and quantified so that the information can be used to improve modeling and simulation. The experiment provides evidence to support the development of closure relations. The simulation includes direct numerical simulation (DNS) or validated CFD results to support the model development when the budget or time frame limits full-scale experiments. The existing models are compact forms of data from past researches and experiments. They are used under appropriate conditions, and often serve as first estimates when new observations and directly relevant data are not available.

**Method as a Service (MaaS).** The MaaS module is emphasized by dotted lines to indicate that ML methods can fill the gap between data and thermal fluid models. ML methods are essential for DDM due to its capability to capture trends of data by nonparametric models. However, if the source of data is uncertain, the data-driven model is also uncertain. Data preprocessing is required to check the consistency* between the model and data before training a data-driven model. For instance, when we use a 3D (three-dimensional) simulation to inform a 1D (one-dimensional) model, we should confirm that the spatiotemporal averaging methods for high-resolution data are consistent with the 1D model. After data reprocessing, ML techniques are applied to accomplish data-driven modeling. Eventually, ML-based closure relations are incorporated into system simulation platforms to enhance the predictability of simulation for a newly designed system and system with different coolants or geometries.

**Platform as a Service (PaaS).** Thermal fluid models with distinct hypotheses can be adapted based on each particular condition, and hence minimize the uncertainty for simulation. Each thermal fluid model requires distinct ML-based closures and may need specific numerical schemes for solutions. Therefore, simulation platforms store the thermal-fluid-model hierarchy based on different degrees of averaging, and provide numerical solvers that are validated by mathematicians. The model selection will become application-oriented, and users can deploy a customized system for dynamics analyses by assembling pre-existing components in the model repository. For instance, when liquid and vapor phases are tightly coupled, it is hard to distinguish interfacial details, and a drift-flux model should be used for this condition [6].

*1.2. Machine learning for thermal fluid simulation*

Machine learning (ML) can be used to develop closure models by learning from the available, relevant, and adequately evaluated data† (ARAED) with nonparametric models. While the concept of ML is not new, the past decade has witnessed a significant growth of capability and interest in machine learning thanking advances in algorithms, computing power, affordable memory, and abundance of data. There is a wide range of applications of machine learning in different areas of engineering practice. In a narrow context of the present study, the machine learning is defined as the capability to create effective surrogates for a massive amount of data from measurements and simulations.

---

* Thermal fluid simulations involve conservation equations with various degrees of averaging from the first principle based on distinct hypotheses. The underlying physics of the conservation equations should be consistent with the experiment or simulation where the available, relevant, and adequately evaluated data (ARAED) are obtained.
† In this study, assumption is made that the data required for ML are available, and their relevance and applicability has been assessed.

Fig. 3 depicts a workflow of employing ML for developing thermal fluid closures. The objective is to construct a function to represent the unknown model that correlates inputs and targets. Since the supervised learning [7] is interested, inputs and targets are essential that can be obtained from ARAED. The $X$ denotes the flow feature space as inputs. The $Y$ presents the response space as targets that are associated with flow features. The subscript $k$ denotes the $k^{th}$ measurement at a certain location. After collecting all relevant datasets, ML models ($ML$) are generalized by a set of nonlinear functions with hyperparameters to represent a thermal fluid closure. Based on different ML methods, various algorithms are employed to seek an optimal solution that allows a ML-based model to fit the observed data. Based on distinct learning purposes, Domingos [8] classified ML methods into five tribes including symbolists, evolutionaries, analogizers, connectionists, and Bayesians. Ling & Templeton [9] evaluated the predictability of various ML algorithms for predicting the averaged Navier-Stoke uncertainty in a high Reynolds region.

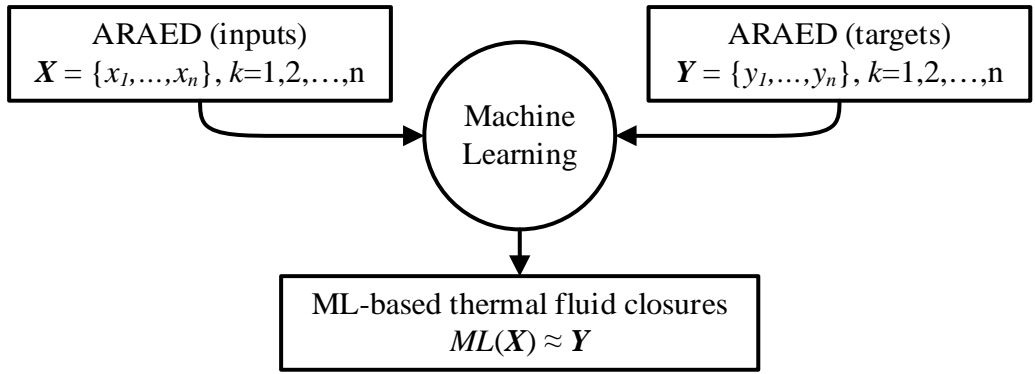

Fig. 3. Workflow of employing ML methods for developing thermal fluid closures.

### 1.3. Thermal fluid data

Fig. 4 provides an overall characterization of thermal fluid data [10] by data type, data source, and data quality. The global data are system conditions and integrated variables such as system pressure, mass flow rate, pressure drop, and total heat input. The local data are time series data at specific locations. The field data are measurements of field variables resolved in space and in time. Traditionally, experiments are a primary source of data, including so-called integral effect tests (IETs) and separate effect tests (SETs). As the name suggests, SETs and IETs are designed to investigate isolated phenomena and complex (tightly coupled) phenomena, respectively. Increasingly, appropriately validated numerical simulations become a credible source of data. This includes high-fidelity numerical simulations (e.g., DNS, and other CFD methods), as well as system-level simulation using computer models in parameter domains that are extensively calibrated and validated. It is noted that datasets vary by their quality regarding the quantity and uncertainty. The amount of data affects the performance of inverse modeling since sufficient data can reduce the model parameter uncertainty in the domain of interest. Within a narrow context of ML for thermal fluid simulation, the data quality can be characterized by the amount of relevant and adequately evaluated data (i.e., data quantity) and associated uncertainty (including measurement uncertainty and other biases, e.g., scaling, processing).

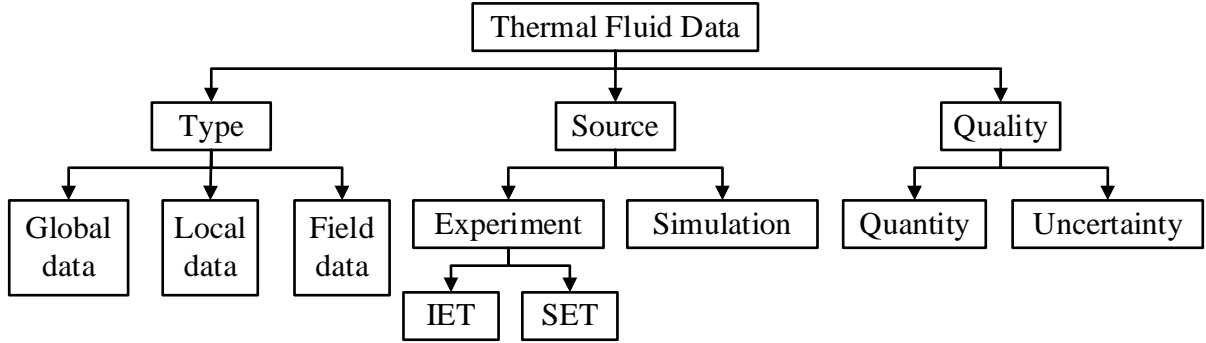

Fig. 4. Hierarchy of thermal fluid data.

### 1.4. Machine learning frameworks for data-driven thermal fluid models

As evident from the preceding discussion, the time is ripe for applying ML methods to assist developments of data-driven thermal fluid models. There is a large variety of ways to embed ML in the models. The three major factors are knowledge of physics (overall model form), available data, and machine learning techniques. It is necessary to incorporate knowledge of underlying physics – whenever such knowledge is available and trustworthy – into ML-based models. The benefits of so-doing are wide-ranging, from preventing models from generating unphysical results to narrowing the search space by reducing the dimension of problems.

In this paper, the focus is placed on the use of neural networks, specifically deep learning (or multilayer neural networks), which has recently emerged as capable and universal approximator. Notably, their hierarchical structures deem appropriate for describing complex models that involve multiple scales.

The objectives of this paper are to develop a system to characterize different approaches to use ML to aid developments of data-driven models in thermal-fluid simulation to help navigate an inter-disciplinary domain (of thermal-fluid simulation, data-driven modeling, and deep learning). The technical approach stems from literature analysis, implementation, and investigation of major types of ML frameworks on synthetic examples. The paper is structured to include a brief account of most relevant works (Section 2), introduction of classification (Section 3), illustrations on heat conduction (Section 4), turbulent flow (Appendix A), and two-phase flow (Section 5), and concluding remarks (Section 6).

## 2. Contemporary work of using ML methodologies in thermal fluid simulation

Insofar, based on a best-effort review of the literature, contemporary work can be grouped into two distinct strategies for employing ML in the field of thermal fluid simulation, namely in algorithm implementation and physics identification. The first approach assumes the physics of fluids is known and applies ML to improve solution performance. For example, Tompson *et al.* [11] utilized convolutional neural networks (CNNs) to accelerate Eulerian fluid simulations. Ladický *et al.* [12] applied regression forests to accelerate smoothed particle hydrodynamics simulations. The other strategy employs ML methodologies to a body of data to capture underlying correlations including recognizing governing equations. Brunton, Proctor & Kutz [13] utilized sparse regression with time-series data to recover Navier-Stokes equations. Not in the thermal fluid domain, but the relevance is the work of Mills, Spanner & Tamblyn [14]. They showed that CNNs with millions of training data recovered the effective form of the Schrödinger equation.

Aside from extracting governing equations from data, ML has been applied to construct surrogates of closure relations. Limited work in this direction relies on supervised learning with the training data from either DNS or large eddy simulation (LES). The applications include both single-phase and two-phase flow problems. Ma, Lu & Tryggvason [15-17] utilized artificial neural networks (ANNs) to obtain closures for stream stress and surface tension force. They implemented ANN-based closures in a two-fluid model for simulation of isothermal bubbly flow in a vertical box channel.

Parish & Duraisamy [18] proposed the field inversion and machine learning (FIML) framework that used a Gaussian process to assimilate data. They demonstrated that source terms of the heat conduction equation could be inferred from data to reconstruct spatial temperature profiles. The FIML was also applied to the $k$-$\omega$ turbulence model [19] for assimilating DNS data to reduce model form errors. The modified turbulence model was used to improve Reynolds-averaged Navier-Stokes (RANS) equations for simulating a single-phase planar channel flow. In a more recent work, Zhang & Duraisamy [20] replaced Gaussian functions by feedforward neural networks (FNNs) to enable spatiotemporal modifications. Tracy, Duraisamy & Alonso [21] utilized an FNN-based closure to learn the results from the Spalart–Allmaras [22] turbulence model.

Wu *et al.* [23] and Wang *et al.* [24, 25] used random forest regression to build a discrepancy field of Reynold stress between DNS and RANS simulation results. Then they used a modified Reynolds stress field to improve the prediction of the RANS model for test flows. Ling, Kurzawski & Templeton [26] trained Reynolds stress closures by tensor basis deep neural networks (DNNs) with DNS and LES data. The inputs of DNNs were the mean strain and rotation rate tensors obtained from RANS simulation using eddy viscosity models. The results indicated that DNNs improved RANS simulation for flows in different geometries with various Reynolds numbers.

The current reference works demonstrate several ways to implement ML-based models in conservation equations. However, those works do not discuss the implementation from perspectives of thermal fluid data that include data type, data source, and data quality. Assumptions behind experiments affect the selection of appropriate frameworks. This work aims at developing a classification system that providing a comprehensive overview of using ML to maximize predictive capabilities of thermal fluid simulation. Such a classification system allows ML frameworks to become transparent regarding assumptions, workflows, and knowledge and data requirements. ML frameworks are solution algorithms to achieve data-driven modeling of thermal fluid simulation. With a classification system, we can select appropriate frameworks based on the available data and knowledge.

## 3. Classification of ML frameworks for thermal fluid simulation

ML strategies employed in thermal fluid simulation include various frameworks to leverage the value of data from experiments or validated simulations such as DNS, LES, or RANS. Several recent studies aim at closing the mass-momentum-energy conservation equation by ML-based closures, while others on extracting governing equations from data. All frameworks have the goal to represent underlying correlations behind data and to capture data in compact forms for simulation. Based on distinct strategies of incorporating ML into thermal fluid simulation, we propose a classification into five frameworks including physics-separated ML (PSML or Type I ML), physics-evaluated ML (PEML or Type II ML), physics-integrated ML (PIML or Type III ML), physics-recovered (PRML or Type IV ML), and physics-discovered ML (PDML or Type V ML).

Type I ML is physics-separated because it requires the separation of scales [27, 28]. Closure models are independently built upon data, and then they are implemented in conservation equations. Type II ML is physics-evaluated. The framework includes simulation based on prior knowledge. When discrepancies occur between observations and simulation, the observed data become references to inform simulation to achieve data-model consistency. Type III ML is physics-integrated since there is no need to separate scales. Instead, ML-based closure models are embedded and trained in conservation equations. Type IV ML is physics-recovered because it aims at recovering the form of governing equations from data. Type V ML is physics-discovered. It is end-to-end ML that ultimately relies on learning algorithms to figure out hidden physics from a considerable amount of data.

ML frameworks are solution algorithms to allow thermal fluid simulation to leverage the value of data. Fig. 5 depicts the hierarchy of ML frameworks based on the goal structuring notation (GSN) [29, 30]. GSN can present the logic of argumentation by a graphic notation. Fig. 6 gives definitions of principal components defined by GSN. Fig. 5 shows that the top goal ($G_{top}$) is to maximize predictive capabilities of thermal fluid simulation by ML. Then there are three sub-goals following by the top goal about how to achieve the top goal. First, ML can assimilate data to construct closure relations. Based on different data sources and assumptions, solution algorithms to the first sub-goal ($G_1$) can be found by Type I, Type II, and Type III ML. Second, since a system can be nonlinear and include multiscale dynamics, we may not want to assume governing equations are known. Instead, we rely on ML to recover the form of equations by assuming that a thermal fluid process can be effectively captured by a PDE model. Type IV ML is the solution to the second sub-goal ($G_2$). Third, if data are immense enough, ML is expected to discover hidden physics directly through data. Type V ML is the solution to the third sub-goal (G3). It is noted that first two sub-goals lead to solutions that converge to conservation equations. Section 3.2-3.6 introduce each framework in detail.

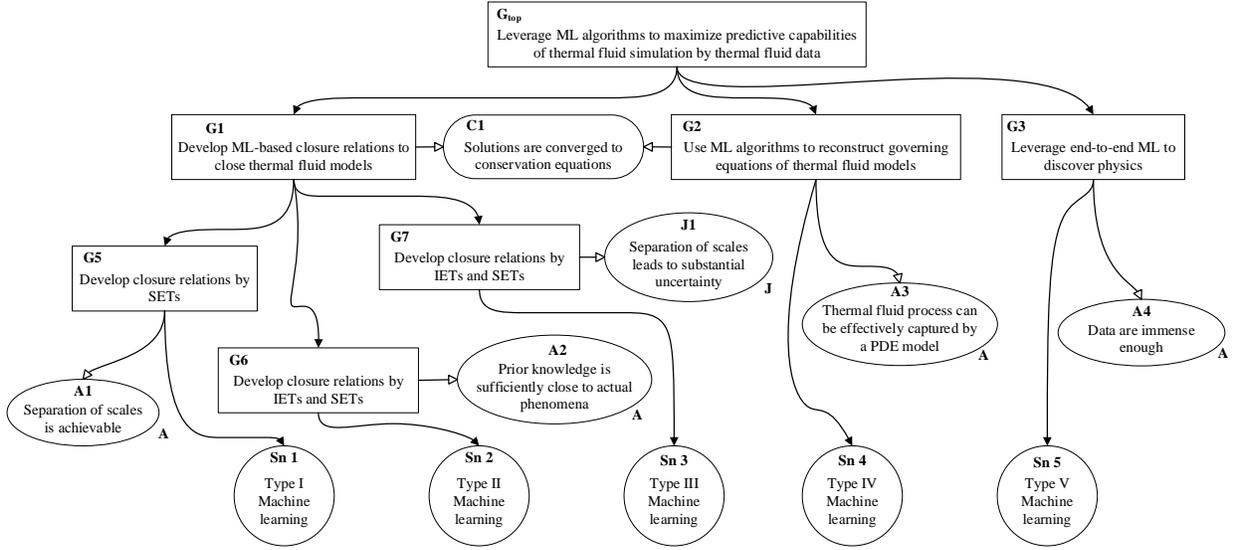

Fig. 5. Hierarchy of machine learning (ML) frameworks for thermal fluid simulation.

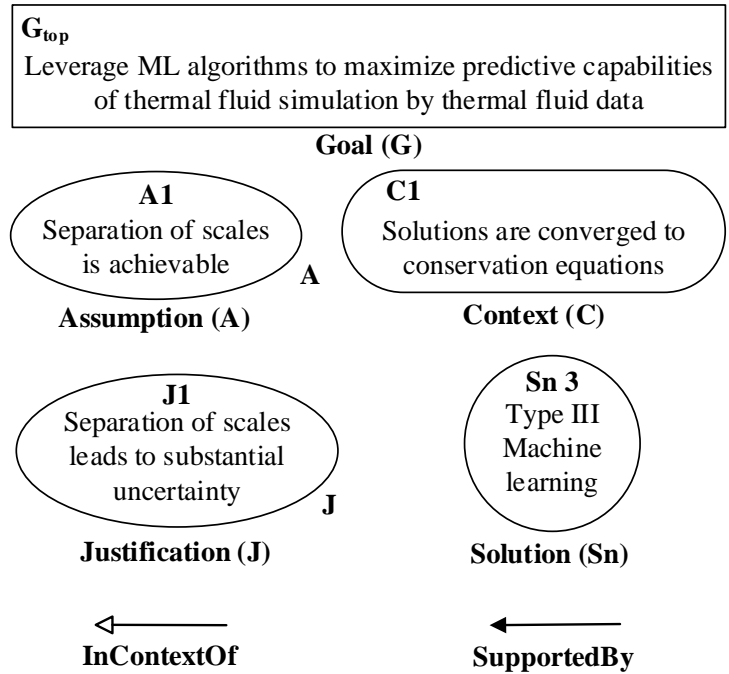

Fig. 6. Principal components of the ML framework hierarchy using the notation by global structuring notation (GSN).

## 3.1. Criteria for classifying ML frameworks for thermal fluid simulation

Each framework has its distinct goal and approach to leverage data. Since we classify five frameworks, we build the classification system based on four conditions. First, we examine whether solutions are converged meaning that solutions conserve the mass-momentum-energy balance in a control volume. Second, we check if the framework focuses on developing fluid closures. Third, we distinguish Type III ML from other frameworks because it inherently ensures data-model consistency. Finally, the last condition is about the separation of scales. Accounting for all four conditions, we categorize five distinct types of ML frameworks for thermal fluid simulation based on the following four criteria:

*Criterion 1: Is PDE involved in thermal fluid simulation?*

The first criterion examines whether conservation equations are involved in thermal fluid simulation. Type V ML relies on ML to discover the underlying physics directly from data and to deliver equivalent surrogates of governing equations. Type V ML is an extreme case when there is no prior knowledge, and we must purely depend on the observed data. By this criterion, we can distinguish Type V ML from other four ML frameworks.

*Criterion 2: Is the form of PDEs given?*

The second criterion inspects if the form of conservation models is known. Type IV ML does not make biases on selecting physics models; instead, it recovers the exact form of conservation models based on data. Therefore, we can distinguish Type IV ML from Type-I, Type II, and Type III ML.

*Criterion 3: Is the PDE involved in the training of closure relations?*

PDEs are involved in Type I, Type II, and Type III ML. Therefore, the goal is to develop closure models in nonparametric forms to close conservation equations. Criterion 3 checks whether conservation equations are involved in the training of ML-based closures. Traditionally, the assumptions [31, 32] of scale separation and physics decomposition are essential to develop closure models. The former allows us to set up SETs for various scales while the latter decomposes closure relations into different physics within the same scale. However, in many thermal fluid processes, the physics (physical mechanisms) is tightly coupled. Type III ML avoids these two assumptions by training closure models that are embedded in PDEs. By this criterion, we can distinguish Type III ML from Type I and Type II ML.

*Criterion 4: Is a scale separation assumption required for the model development?*

This criterion tests whether the model development requires the separation of scales. This hypothesis isolates closure relations from conservation equations so that the models can be separately built and calibrated by SETs. The scale separation is essential for Type I ML because it only relies on data to construct closure models. However, the data by SETs may have been distorted, while IETs are designed to capture (a selected set of) multi-physics phenomena.

Table 1 summarizes the criteria to classify the five distinct types of ML frameworks for thermal fluid simulation.

Table 1. Criteria for the ML framework classification.

| Classification Criteria | Type I ML (PSML) | Type II ML (PEML) | Type III ML (PIML) | Type IV ML (PRML) | Type V ML (PDML) |
|---|---|---|---|---|---|
| 1. Is PDE involved in thermal fluid simulation? | Yes | Yes | Yes | Yes | No |
| 2. Is the form of PDEs given? | Yes | Yes | Yes | No | No |
| 3. Is the PDE involved in the training of closure relations? | No | No | Yes | No | No |
| 4. Is a scale separation assumption required for the model development? | Yes | No | No | No | No |

### 3.2. Type I machine learning, physics-separated machine learning (PSML)

Type I ML or so-called physics-separated ML (PSML) aims at developing closure models by using SET data. Type I ML assumes that conservation equations and closure relations are scale separable, for which the models are local. Type I ML requires a thorough understanding of the system so that SETs can be designed to support model developments. Fig. 7 depicts the hierarchical decomposition of system simulation that allows physics models to be scale separable. The system can be divided into various sub-systems such as a reactor core, steam generator, reactor coolant system, and emergency core cooling system. The foundations of those sub-systems are multiphase models that require closure relations based on sub-grid-scale physics. Fig. 8 illustrates the workflow about how we can obtain closure models to close conservation equations where the models are separately developed by using SET data. Therefore, we can apply ML-based closures to assimilate data to achieve data-driven thermal fluid simulation.

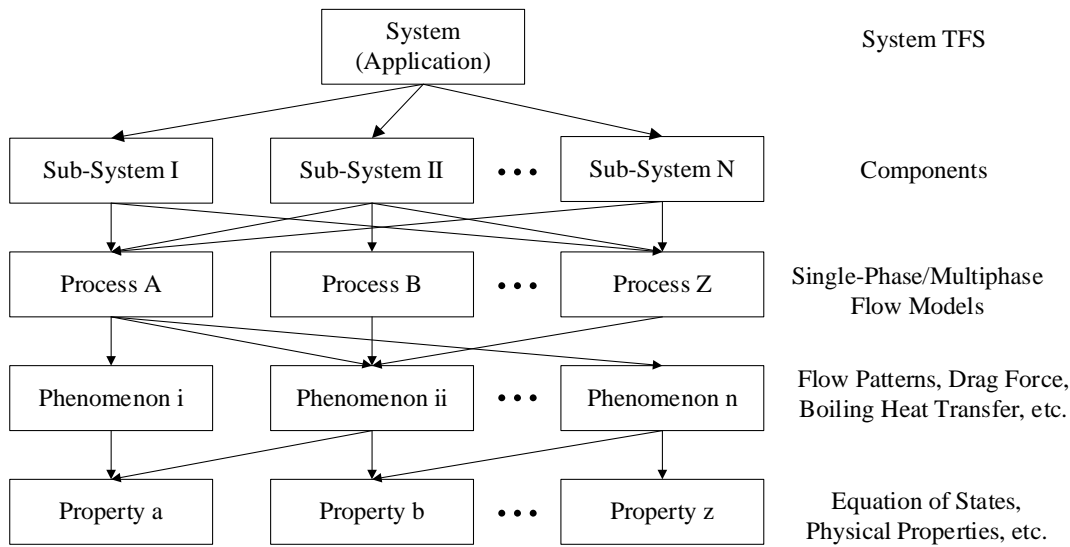

Fig. 7. Hierarchical decomposition of system thermal-hydraulics.

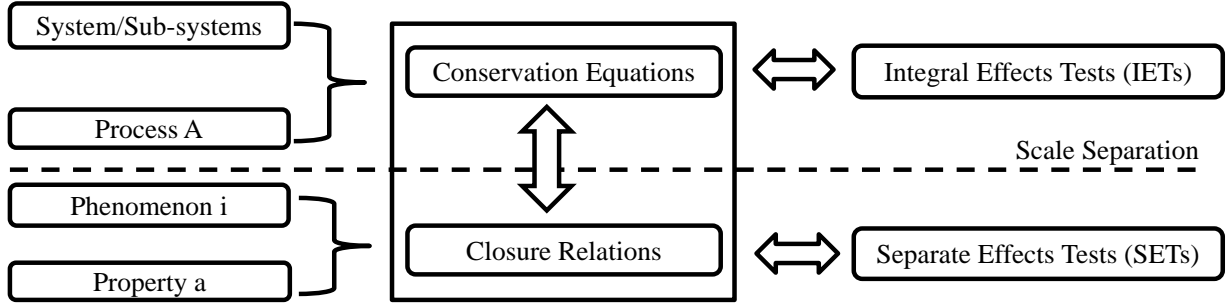

Fig. 8. Based on a scale separation assumption, closure relations can be obtained by SETs to close conservation equations that require IET data for calibration.

Fig. 9 depicts the architecture of Type I ML framework, and it is forward data-driven modeling. The procedure includes the following elements:

*Element 1*. Assume a scale separation is achievable such that closure models can be built from SETs. From either high-fidelity simulations or experiments, collect training data, ($x_k$, $y_k$).

*Element 2*. Preprocess data from element 1 to ensure that data from multi-sources have the same dimension and manipulation such as the selection of averaging methods. Additionally, consider normalizing data so that we can approximately equalize the importance for each data source. For large datasets, employ principal component analysis [33] can be helpful to reduce the dimension of data.

*Element 3*. Compute flow features or system characteristics, $X$, as training inputs for element 5.

*Element 4*. Calculate the corresponding outputs ($Y$) of the desired closures from data as training targets that can supervise ML algorithms to learn from data.

*Element 5*. Utilize ML algorithms to build a correlation between inputs and targets. After the training, output the ML-based closure model, $ML(X)$, to element 6.

*Element 6*. Constrain the ML-based closure, $g(ML(X))$, to satisfy model assumptions and to ensure the smoothness of model outputs since it needs to be solved with PDEs. It is noted that this element is not essential if assumptions are not applicable.

*Element 7*. Implement the ML-based closure into conservation equations, and solve PDEs for predictions with the embedded ML-based closure that is iteratively queried.

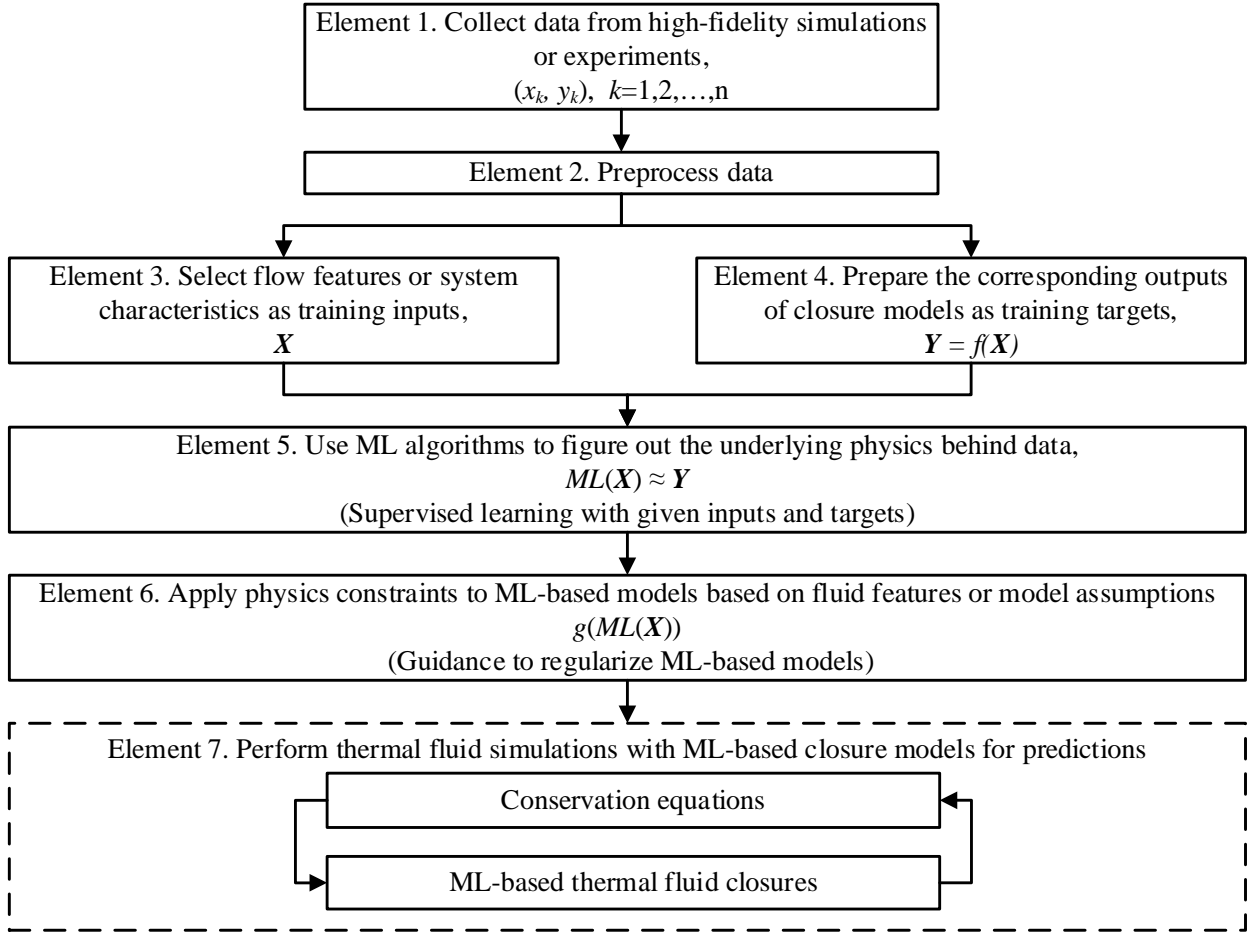

Fig. 9. Overview of Type I ML framework with a scale separation assumption.

Type I ML satisfies the criteria from Table 1 except the third criterion. The quality of SET data largely controls the performance of closure models obtained by Type I ML. While the experimental uncertainty in each SET may be controlled and reduced, the process uncertainty (dominated by design assumptions) is irreducible. We refer that PDEs and closure relations are decoupled in Type I ML. It can cause model biases between conservation equations and closure relations. It is noted that inferring model parameters from data belong to inverse problems which are ill-posed [34]. For ML models, a small change in inputs can result in large uncertainty in outputs. While implementing ML-based closures in PDEs, the uncertainty can lead to a discontinuity that fails numerical simulation. For more practices related to Type I ML, readers are referred to Ma *et al.* [15-17], Parish & Duraisamy [18], Zhang & Duraisamy [20], Tracy *et al.* [21, 35], Singh & Duraisamy [36], and Chang & Dinh [37, 38].

### 3.3. Type II machine learning, physics-evaluated machine learning (PEML)

Type II ML or so-called physics-evaluated machine learning (PEML) focuses on reducing the uncertainty for conservation equations. It requires prior knowledge on selecting closure models to predict thermal fluid behaviors. Type II ML utilizes high-fidelity data to inform low-fidelity simulation. Comparing to high-fidelity models, ROMs can efficiently solve engineering design problems within an affordable time frame. However, ROMs may produce significant uncertainty in predictions. Type II ML can improve the uncertainty of low-fidelity simulation by reference data. Since the physics of thermal fluids is nonlinear, ML algorithms are employed to capture the underlying correlation behind high-dimensional data. The framework requires training inputs such as flow features that represent the mean flow properties. Training targets are the responses that correspond to input flow features.

Fig. 10 depicts the framework of Type II ML, and it includes the following procedures:

*Element 1*. Perform low-fidelity simulation ($\boldsymbol{\Psi_L}$) to generate data for calculating input flow features.

*Element 2*. Perform high-fidelity simulation ($\boldsymbol{\Psi_H}$) with identical system characteristics in element 1. High-fidelity data are used to compute targets in element 5.

*Element 3*. Average high-fidelity data to match the dimension of low-fidelity data. The averaging method should preserve the consistency between high-fidelity and low-fidelity simulations. Additionally, normalizing data can accelerate the training of ML. For large datasets, principal component analysis [33] can reduce the dimension of data.

*Element 4*. Calculate flow features, $X(\boldsymbol{\Psi_L})$, as training inputs to element 6.

*Element 5*. Compute targets, $f(X(\boldsymbol{\Psi_H}))$, as the responses to input flow features by high-fidelity data. Targets can also be discrepancy/error, $\varepsilon(\boldsymbol{\Psi_H}, \boldsymbol{\Psi_L})$, between high-fidelity and low-fidelity data.

*Element 6*. Use an ML algorithm to represent the underlying correlation between flow features and discrepancy/error of flow properties. After the training, output an ML-based discrepancy/error model, $ML(X(\boldsymbol{\Psi_L}))$, to element 8.

*Element 7*. Execute new low-fidelity simulation ($\boldsymbol{\Psi'_L}$) under predicting conditions. Then use the solution to obtain flow features as inputs to element 8.

*Element 8*. Use flow features from element 7 as inputs to query values from the ML-based model, $ML(X(\boldsymbol{\Psi'_L}))$. Output values of a fluid closure in a fixed field to element 9.

*Element 9*. Implement the results from element 8 in the low-fidelity model ($\boldsymbol{\Psi_L}$) for predictions.

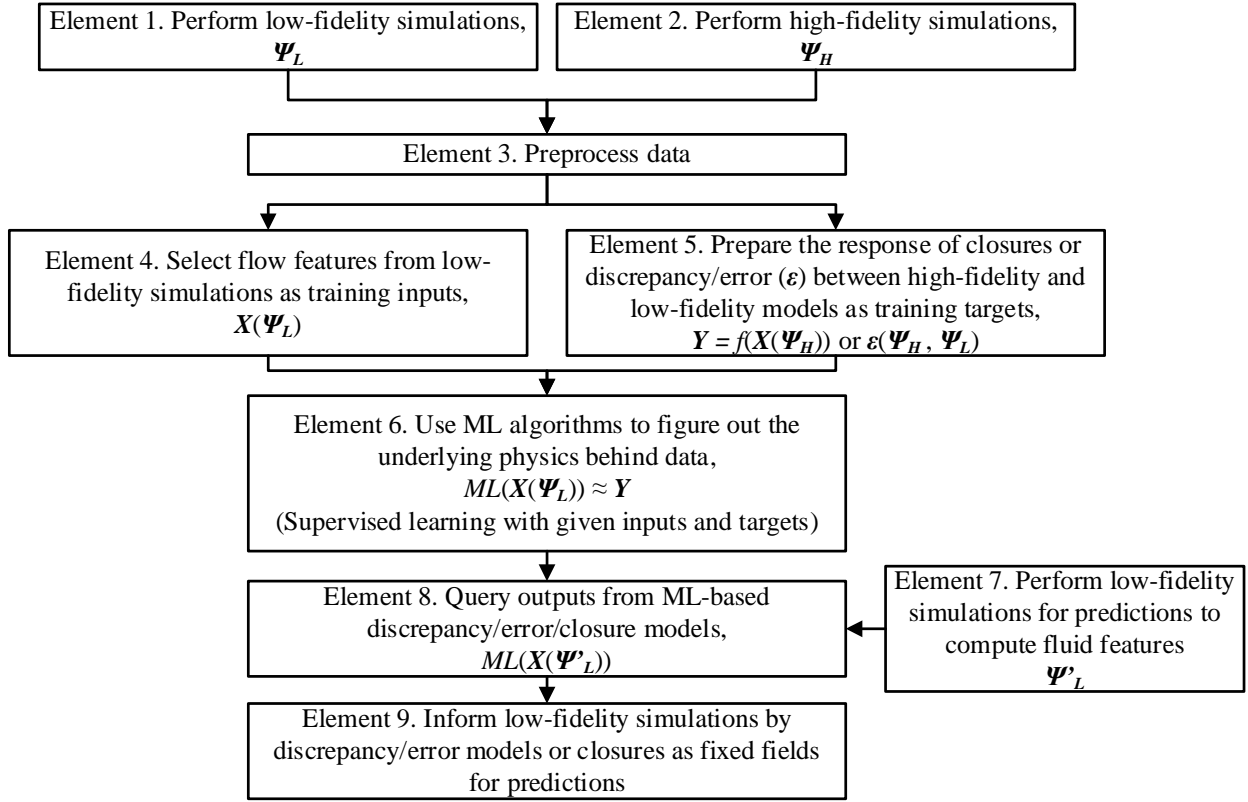

Fig. 10. Overview of Type II ML framework.

Type II ML satisfies the first two criteria in Table 1. We refer that PDEs and closure relations are loosely coupled in Type II ML because PDEs are only used for calculating input flow features. The framework provides a one-step solution to improve low-fidelity simulation. Model uncertainty is not accumulated in Type II ML because numerical solvers do not interact with ML models. However, Type II ML exists an open question about what the magnitude of initial errors can be before it is too late to bring a prior solution to a reference solution. For more detailed examples of Type II ML, readers are referred to Ling & Templeton [9], Ling, *et al.* [39], Wu *et al.* [23], Wang *et al.* [24, 25], Ling *et al.* [26], and Zhu & Dinh [40].

### 3.4. *Type III machine learning, physics-integrated machine learning (PIML)*

To the best knowledge of the authors, Type III ML or so-called physics-integrated ML (PIML) is introduced and developed for the first time in this work. Type III ML aims at developing closure relations to close thermal fluid models without a scale separation assumption. Closure models are embedded and trained in system dynamics. Training data can be obtained from SETs and IETs. Notably, Type III ML can lead the paradigm shift of using ML in thermal fluid simulation because it allows the direct use of field data from IETs. Fig. 11 shows the framework of Type III ML.

Inputs for Type III ML do not directly come from observations; instead, they are solutions of PDEs. Type III ML involves the following elements:

*Element 1*. Collect data, $(x_k, y_k)$, from high-fidelity simulations or experiments that are used to compute targets for the training.

*Element 2*. Preprocess the data from element 1 to ensure that data from multi-sources are consistent with conservation equations regarding the dimension and manipulation such as the selection of averaging methods. Additionally, consider normalizing data so that we can approximately equalize the importance for each data source. For large datasets, employ principal component analysis [33] can reduce the dimension of data.

*Element 3*. Prepare training targets ($Y$) from data that corresponds to PDE solutions.

*Element 4*. For the initial step of the sub-framework, calculate flow features ($X$) from data as training inputs for element 5. After that, flow features are computed based on PDE solutions from element 6.

*Element 5*. Adjust model parameters of an ML-based closure, $ML(X)$, using an ML algorithm. Then output the ML-based closure to element 6.

*Element 6*. Solve conservation equations with the ML-based closure that is iteratively queried during a solution scheme.

*Element 7*. Check if the solution from element 6 converges to the target values within a tolerance interval. If the convergence test does not pass, go to element 4 and continue the loop in the sub-framework. If the result is converged, output the conservation model with the embedded ML-based closure to element 8. The selection of tolerance intervals is case-dependent.

*Element 8*. Solve the PDE model from element 7 with new system characteristics for predictions.

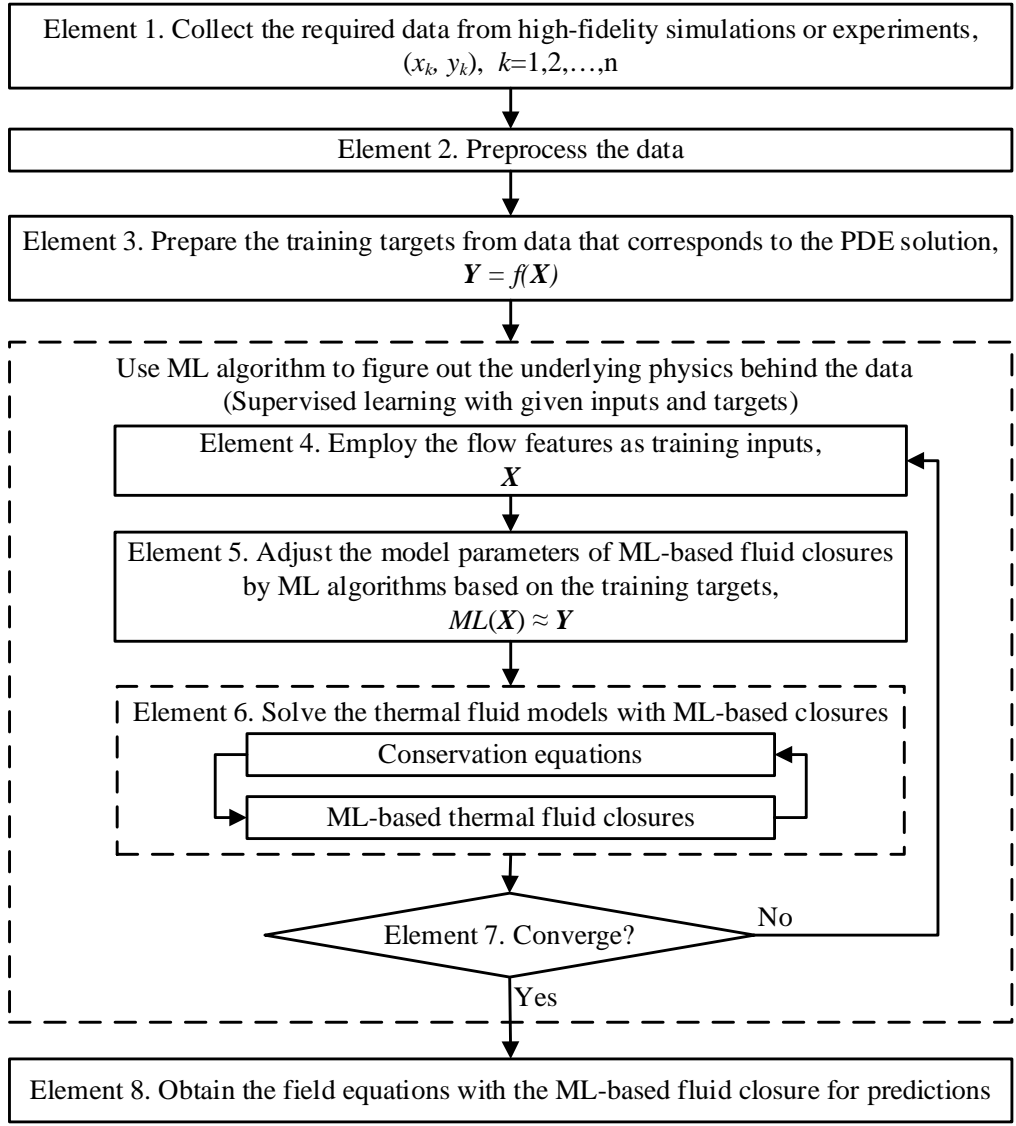

Fig. 11. Overview of Type III ML framework.

Type III ML satisfies most criteria in Table 1 except for the fourth criterion. We refer that PDEs and closure relations are tightly coupled in Type III ML. It is a challenging problem. Such tightly coupled multiscale problems require that numerical solutions (of the governing PDE system) are realized (hence evolving datasets for training) whenever ML algorithms tune model parameters. Therefore, Type III ML is computationally expensive. The research on Type III ML methodology promises a high-potential impact in complex thermal fluid problems where the separation of scales or physics decomposition may involve significant errors.

### 3.5. Type IV machine learning, physics-recovered machine learning (PRML)

Type IV ML or so-called physics-recovered ML (PRML) aims at recovering the exact form of PDEs. Fig. 12 depicts the framework of Type IV ML. It requires no assumption about the form of governing equations. Instead, the framework requires to construct a candidate library that includes components of governing equations such as time derivative, advection, diffusion, and higher order terms. The procedure contains the following elements:

*Element 1*. Collect time series data ($\omega_t$) from either validated simulations or experiments.

*Element 2*. Build a library, $\Theta(X)$, for candidate terms ($X$) in governing equations.

*Element 3*. Reconstruct governing equations using the time derivative term ($X_t$) and the optimal combination of candidate terms ($X$) by sparse regression [13] with a sparse vector ($\xi$) that follows Occam's razor [41].

*Element 4*. Solve the recovered governing equation with new system characteristics for predictions.

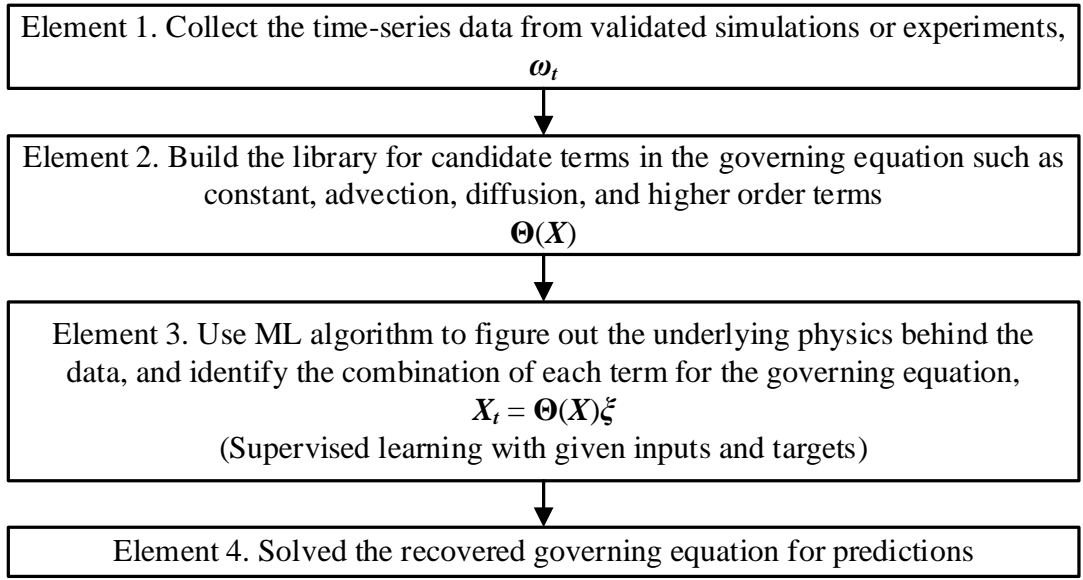

Fig. 12. Overview of Type IV ML framework.

Type IV ML only satisfies the first criterion in Table 1. The challenge of Type IV ML can be the recovery of closure relations in thermal fluid models. Closure models are usually complex, and they are hard to be represented by each derivative term. Therefore, it is an open question about how to apply Type IV ML for complex flow system such as turbulence modeling. For more practices related to Type IV ML, readers are referred to Brunton *et al.* [13].

### 3.6. Type V machine learning, physics-discovered machine learning (PDML)

Type V ML or so-called physics-discovered ML (PDML) is the extreme case. Type V ML is used for either condition. First, it assumes no prior knowledge of physics. Second, it assumes existing models and modeling tools are not trustworthy or not applicable for thermal fluid systems under consideration. More generally, Type V ML is "equation-free" and instrumental in the search for a new modeling paradigm for complex thermal-fluid systems. Type V ML does not involve conservation equations nor satisfy any criterion in Table 1. Instead, it wholly relies on data to discover the effective predictive models. However, such situation rarely occurs because there are usually physics principles or hypotheses that can be postulated to reduce the dimension of problems. For the discussion related to Type V ML, readers are referred to Mills *et al.* [14] and Hanna *et al.* [42].

### 3.7. Knowledge and data requirements for ML frameworks in thermal fluid simulation

In the present context of ML, knowledge refers to a body of theoretical and empirical evidence that is available and trustworthy for understanding and description of physical mechanisms that underlie thermal fluid processes under consideration. This knowledge can guide selecting model forms, including conservation equations and corresponding closure relations, designing experiments, and performing high-fidelity simulations. The data requirements refer to characteristics of the body of data (e.g., types, amount, quality) needed to enable thermal fluid simulation with the required accuracy. In other words, the required data must be sufficient to complement the "knowledge" for building closure models and recovering/discovering the physics.

The form of PDEs are known for Type I, Type II, Type III ML, and the focus is to build closure relations. In traditional modeling approaches, closure models are local, relating a group of (local) source terms (i.e., sub-grid-scale interactions) to a group of (local) flow features. Even when in engineering literature, source terms are expressed regarding global parameters (like flow rate, system pressure), they are used as surrogates for local-valued parameters (through the assumptions that equate global and local conditions).

Type I ML build closure relations independently from PDEs, but it requires a thorough or assumed understanding of the physics that is essential to set up SETs for acquiring data. Globally measured data or locally measured data (using point instruments) are very small amount of data. In such case, complicated ML-based closures are not necessarily the best choice. Therefore, among the frameworks, Type I ML exhibits a minimal data requirement with a maximal knowledge requirement.

Type-II ML assumes prior knowledge of physics that guide the selection of closure relations for thermal fluid simulation. However, the use of prior models yields uncertainty in thermal fluid analyses. This uncertainty (or error) can be inferred by comparing the model prediction to reference solutions from high-fidelity simulations, high-resolution experiments as well as data obtained in IETs that include multi-physics phenomena. Correspondingly, Type II ML requires larger data quantities but less knowledge than Type I ML.

Type III ML trains closure relations that are embedded in conservation equations without invoking a scale separation assumption. IET data can be directly adapted into simulation by applying Type III ML. While the term ML is broad, in the present work ML refers to the use of non-parametric models or even narrower, use of DNNs. This means no prior knowledge of model forms of closure relations. Thus, Type III ML requires less knowledge than Type II ML (which "best-estimated" closure models on the basis of past data). Consequently, Type III ML requires a large body of data to represent models than that of Type II ML.

Type IV ML intends not to make any bias on selecting conservation equations; instead, it recovers the exact PDE form from data. It assumes less prior knowledge but requires more extensive training data than the previous three frameworks.

Type V ML is an extreme case that makes no assumption about prior knowledge or reference solutions for thermal fluid systems under consideration. The aim is to apply ML methods to learn from data, and to establish a data-driven predictive capability. For thermal fluid simulation, it means discovering the effective model form of conservation equations and closure relations. Accordingly, among the frameworks, Type V ML is the most stringent with respect to data requirements (types, quantity, and quality).

Fig. 13 depicts the domain of ML frameworks regarding prior knowledge and data requirements.

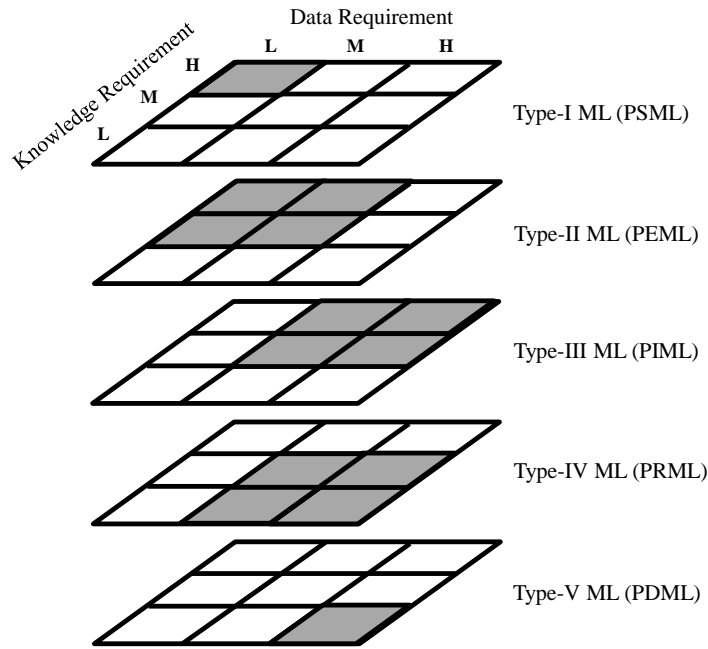

Fig. 13. The domain of various ML frameworks where **L**, **M**, and **H** denote low, medium, and high.

## 4. Heat conduction case study by Type I, Type II, Type III, and Type V ML frameworks

The heat conduction case study is formulated to demonstrate how to employ Type I, Type II, and Type III ML to build ML-based thermal conductivity and to compare results by each framework. Chanda *et al.* used ANN with genetic algorithm [43] to solve inverse modeling for heat conduction problems. In this work, deep learning (DL) [44] is selected as the ML methodology in this task. Principally, any neural network (NN) with more than two layers (one hidden layer with an output layer) is considered as to be DL [45]. Hornik [46] proved that multilayer NNs are universal approximators, and it can capture the properties of any measurable information. This capability makes DL attractive for the closure development in thermal fluid simulation. Notably, we implement NN-based thermal conductivity by FNNs and convolutional neural networks (CNNs) to evaluate the performance of closure relations by distinct NNs.

### 4.1. Problem formulation

We formulate the synthetic task using a 2D (two-dimensional) heat conduction model given by Eq. (1) where $k(T)$ is nonlinear thermal conductivity. To generate training data, Eq. (2) shows a temperature-dependent model for $k(T)$ where $c$, $\sigma$, and $\mu$ are constant parameters. Table 2 gives two parameter sets (baseline and prior sets) to generate data. While demonstrating ML frameworks, $k(T)$ becomes NN-based thermal conductivity.

$$\frac{\partial}{\partial x}\left[k(T)\frac{\partial T}{\partial x}\right] + \frac{\partial}{\partial y}\left[k(T)\frac{\partial T}{\partial y}\right] = 0 \qquad (1)$$

$$k\left(T\right) = \frac{c}{\sigma\sqrt{2\pi}}e^{-\frac{(T-\mu)^2}{2\sigma^2}} \qquad (2)$$

Table 2. Two parameter sets for the thermal conductivity model.

| Dataset | $c$ (W/m) | $\sigma$ (K) | $\mu$ (K) |
|---|---|---|---|
| Baseline set for producing synthetic data | $7.2 \times 10^4$ | 300 | 1200 |
| Prior set for producing inputs required by Type II ML | $7.2 \times 10^4$ | 600 | 2100 |

Two numerical experiments are designed to emulate IETs and SETs for manufacturing synthetic data by solving Eq. (1) using parameters sets in Table 2. IETs provide field data, for instance, 2D temperature fields by an infrared camera. SETs offer global data such as a 1D measurement by thermocouples. Synthetic data are used for training and validating NN-based thermal conductivity. Type I ML can only use SET data because of a scale separation assumption. Type II ML can only use SET data because the goal is to improve the prior thermal conductivity by the baseline. Type

III and Type V ML use field data. We compare Type I and Type II ML using training data from SETs. Then Type III and Type V ML are compared by field data from IETs.

## 4.2. Manufacturing synthetic data for ML frameworks

Numerical solutions assumes piecewise-linear temperature between grids [47] with Dirichlet type boundaries. We further assume conductivity profiles are also piecewise-linear between mesh points.

### 4.2.1. Manufacturing IET data

IETs are measurements of temperature fields. Synthetic IET data are generated by Eq. (1) with the baseline set in Table 2. Fig. 14 illustrates the layout of IET experiments with four constant boundary temperatures. We change $T_{west}$ for various observations and fix the boundary temperature (1300K) at the east side. The $T_{north}$ and $T_{south}$ are linearly dependent on the west boundary condition. We prepare three training datasets by including distinct data quantities and three validating datasets by changing $T_{west}$. Table 3 gives the metadata of each training or validating dataset. All observations are uniformly sampled within a given temperature range.

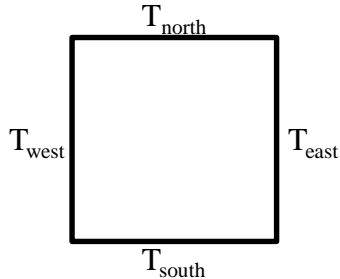

Fig. 14. Schematic of integral effects tests (IETs) for measuring temperature fields.

Table 3. Summary of IET training and validating datasets.

| Dataset | Data Quantity | Temperature Range at $T_{west}$ | Description |
|---------|---------------|----------------------------------|-------------|
| T1 | 11 observations | [1000K, 1100K] | Training dataset |
| T2 | 100 observations | [1000K, 1100K] | Training dataset |
| T3 | 1000 observations | [1000K, 1100K] | Training dataset |
| P1 | 1000 observations | [1000K, 1100K] | Validating dataset |
| P2 | 1000 observations | [900K, 1000K] | Validating dataset |
| P3 | 1000 observations | [800K, 900K] | Validating dataset |

### 4.2.2. Manufacturing SET data

SETs are global measurements by thermocouples. Fig. 15 depicts the layout of SETs for obtaining mean temperature and heat conductivity data. A heater is on top of the sample to maintain a constant temperature ($T_H$). Thermal insulations are installed on the outside surface. The coolant at the bottom removes the heat with a constant heat transfer coefficient. Eq. (3) calculates temperature profiles within the sample using parameter sets in Table 2. Eq. (4) calculates the observed heat conductivity ($k_{obs}$), and the mean temperature is obtained by arithmetic averaging $T_H$ and $T_C$.

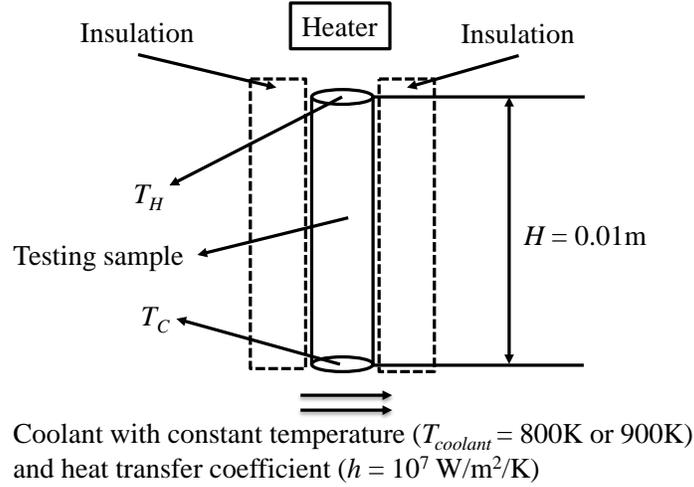

Coolant with constant temperature ($T_{coolant} = 800$K or 900K) and heat transfer coefficient ($h = 10^7$ W/m²/K)

Fig. 15. Schematic of separate effects tests (SETs) for measuring thermal conductivity as the function of sample's mean temperature.

$$\frac{\partial}{\partial x}\left[ k(T)\frac{\partial T}{\partial x} \right] = 0 \tag{3}$$

$$k_{obs}\frac{T_H - T_C}{H} = h(T_C - T_{coolant}) \tag{4}$$

We generate two training datasets with two coolant temperatures to explore the effect by different data qualities. Table 4 shows the metadata of SET datasets. A large temperature gradient across the testing sample increases the nonlinearity of temperature profiles. For each training set, we uniformly sample 41 $T_H$ from Eq. (5) to keep mean temperatures in SETs within the same range as IETs.

Table 4. Summary of SET training datasets.

| Dataset | Data Quantity | Data Quality | $T_{coolant}$ (K) | Description |
|---------|---------------|--------------|-------------------|-------------|
| S1 | 41 observations | Low | 800 | Training dataset |
| S2 | 41 observations | High | 900 | Training dataset |

$$\left( T_{H,\max}, T_{H,\min} \right) = \left( 2T_{IET,\max} - T_{coolant}, 2T_{IET,\min} - T_{coolant} \right) \tag{5}$$

### 4.3. Implementation

#### 4.3.1. Implementation of the heat conduction task by Type I, Type II, Type III, and Type V ML frameworks

We present Type I ML in Algorithm 1. SET data are generated by Eq. (3) with the baseline set in Table 2. Inputs and targets are temperatures and thermal conductivities. After the training, FNN-based thermal conductivity is implemented in Eq. (1) for predictions.

---
**Algorithm 1 Type I ML for 2D heat conduction problem with Dirichlet BC.**

---
**Input:** Training inputs ($T_{baseline}$, *element 3* in Fig. 9) and training targets ($k_{baseline}$, *element 4* in Fig. 9) from SETs (*element 1* in Fig. 9)
**Output:** Temperature fields for predictions (*element 7* in Fig. 9)
1: **for** all epochs < maximum_epoch **do** (*element 5* in Fig. 9)
2:     // Build a conductivity model using FNNs
3:     $k(T) \leftarrow FNN(T)$
4:     **for** all inputs $(T_{baseline}, k_{baseline}) \in$ training datasets **do**
5:         Update hyperparameters for each layer in FNNs
6: Implement $k(T)$ into Eq. (1) (*element 7* in Fig. 9)
7: Solve Eq. (1) with Dirichlet boundaries for predictions (*element 7* in Fig. 9)

---

Algorithm 2 outlines the application of Type II ML. SET data are generated by Eq. (3) with prior and baseline sets in Table 2. Targets are thermal conductivities calculated by the baseline set. However, inputs are mean temperatures by the prior set. After the training, FNN-based thermal conductivity can be queried by new input temperature fields from solutions with new boundary conditions. Once thermal conductivities are obtained, they are implemented as fixed fields into Eq. (1) to obtain temperature profiles for predictions. Therefore, Type II ML first solves the heat conduction equation with an initially guessed parameter set, and then improve the solution by the FNN-based closure.

**Algorithm 2 Type II ML for 2D heat conduction problem with Dirichlet BC.**

**Input:** Training inputs ($T_{prior}$) (*element 4* in Fig. 10) and training targets ($k_{baseline}$, *element 5* in Fig. 10) from SETs (*element 2* in Fig. 10)

**Output:** Temperature fields for predictions (*element 10* in Fig. 10)

  1: Solve Eq. (3) with the prior set in Table 2 (*element 1* in Fig. 10)

  2: Use temperature profiles ($T_{prior}$) as inputs for training NNs (*element 4* in Fig. 10)

  3. **for** all epochs < maximum_epoch **do** (*element 6* in Fig. 10)

  4:    // Build surrogates for thermal conductivities using FNNs

  5:    $k(T) \leftarrow FNN(T)$

  6:    **for** all inputs $(T_{prior}, k_{baseline}) \in$ training datasets **do**

  7:      Update weights and biases for each layer in FNNs

  8: Solve Eq. (1) again with the prior set in Table 2 and new boundaries (*element 7* in Fig. 10)

  9: Query $k(T'_{prior})$ by new temperature fields ($T'_{prior}$) (*element 8* in Fig. 10)

10: Implement the thermal conductivities as fixed fields into Eq. (1)**,** and solve the equation to

11: obtain temperature profiles for predictions (*element 9* in Fig. 10)

Algorithm 3 outlines the procedure of Type III ML. IET field data are generated by Eq. (1) with the baseline set in Table 2. Targets are baseline temperature fields. Inputs are the solution by Eq. (1) with NN-based thermal conductivity. During the training, inputs are iteratively updated due to the change of weights and biases in NNs. After the training, the model is ready for predictions. Type III ML ensures data-model consistency because PDEs are involved in the training.

**Algorithm 3 Type III ML for 2D heat conduction problem with Dirichlet BC.**

**Input:** Training targets ($T_{baseline}$, *element 3* in Fig. 11) and training inputs ($T_{PDE}$, *element 4* in Fig. 11)

**Output:** Temperature fields for predictions

1: **for** all outer < maximum_outer_iteration **do**

2:    **for** all epoch < maximum_epoch **do** (*element 5* in Fig. 11)

3:      // Build conductivity models using both FNNs and CNNs

4:      $k(T) \leftarrow FNN(T), \; CNN(T)$

5:      **for** all inputs $T_{PDE} \in$ PDE solutions **do**

6:        Update hyperparameters for each layer in FNNs and CNNs

7: Solve Eq. (1) using NN-based thermal conductivity for obtaining new $T_{PDE}$ as new training

8: inputs for NNs (*element 6* in Fig. 11)

Algorithm 4 shows the implementation of Type V ML using IET data. The data are generated by Eq. (1) with baseline and prior sets in Table 2. Inputs are temperate profiles by the prior set. Outputs are the discrepancies ($\delta T$) between prior and baseline temperature fields. After the training, discrepancy fields are queried by new temperature fields with the prior set and new boundary conditions. Then the improved temperature fields are obtained by adding discrepancy fields into prior temperature fields. Therefore, the predicted temperature fields by Type V ML are not constrained by the governing equation.

| **Algorithm 4 Type V ML for 2D heat conduction problem with Dirichlet BC.** |
|---|

**Input:** baseline temperature ($T_{baseline}$) and training inputs ($T_{prior}$)
**Output:** Temperature fields for predictions
1: Solve Eq. (1) with the prior set in Table 2
2: Use temperature profiles ($T_{prior}$) as inputs for training NNs
3: Compute discrepancies between baseline temperature fields ($T_{baseline}$) and prior temperature
4: fields ($T_{prior}$)
5:     $\delta T = T_{baseline} - T_{prior}$
6: **for** all epochs < maximum_epoch **do**
7:     // Build surrogates for thermal conductivities using FNNs
8:     $\delta T \leftarrow FNN(T_{prior})$
9:     **for** all inputs $(T_{prior}, \delta T) \in$ training datasets **do**
10:         Update weights and biases for each layer in FNNs
11: Solve Eq. (1) again with the prior set in Table 2 and new boundaries
12: Query $\delta T(T'_{prior})$ by new temperature fields $(T'_{prior})$
13: Improve temperature profiles for predictions by $\delta T(T'_{prior})$

Eq. (6) defines the root-mean-square error (RMSE) to evaluate the performance of each ML framework where $i$ denotes the $i^{th}$ observation, $N$ is the total data points in the $i^{th}$ observation, and $j$ presents the $j^{th}$ solution. For each validating dataset, we calculate mean RMSE by using arithmetic averaging.

$$RMSE_i = \sqrt{\frac{\sum_j \left(T_{Model,j} - T_{data,j}\right)^2}{N_i}}$$
(6)

*4.3.2. Implementation of NN-based thermal conductivity model*

*4.3.2.1. FNN-based thermal conductivity model*

We use FNNs and CNNs to reconstruct thermal conductivity from training data. Eq. (7) gives the formulation of FNN-based thermal conductivity where $T$, **x**, and $i$ are the temperature, input vector, and $i^{th}$ training input. The sigmoidal activation function, $1/(1+e^{-x})$, is selected. Eq. (8) shows the structure of the first hidden layer (HL) where *j* is the j<sup>th</sup> hidden units (HUs) and $N_{in}$ is the total inputs from the input layer. The weight (*w*) and bias (*b*) are parameters to be learned based on training data. Eq. (9) presents the general structure of HLs where *k* and $N_{HUk}$ are the layer number and total number of HUs in the k<sup>th</sup> layer. Starting from the second HL, the number of inputs depends on the quantity of HUs from the previous HL. Eq. (10) shows the output layer as a linear combination of HUs from the last HL where *L* is the total layer number of FNNs. For this demonstration, we use three-layer FNNs with ten HUs in each HL, and we fix this structure for all types of ML learning frameworks.

$$k_{DNN} = FNN(\mathbf{x}), with \quad x_i = \left( T_i \right) \tag{7}$$

$$HU_{1j}\left( \mathbf{x} \right) = sigmoid\left( \sum_{i=1}^{N_{in}} w_{1ji}x_i + b_{1j} \right) \tag{8}$$

$$HU_{kj}\left( \mathbf{x} \right) = sigmoid\left( \sum_{i=1}^{N_{HU_k}} w_{kji}HU_{k-1,i} + b_{kj} \right) \tag{9}$$

$$FNN\left( \mathbf{x} \right) = \sum_{i=1}^{N_{HU_{L-1}}} w_{o,i}HU_{L-1,i} + b_0 \tag{10}$$

### 4.3.2.2. CNN-based thermal conductivity model

Fig. 16 depicts the architecture [48] of CNN-based thermal conductivity that includes three convolutional layers and three fully connected layers. We use the ReLU [49] activation for layers in CNNs to accelerate the training. Inputs are temperature fields. After the first convolutional layer, eight feature maps are generated, and each feature map detects the patterns from temperature fields. The second convolutional layer takes inputs from the previous layer, and it outputs 12 feature maps. The third convolutional layer receives inputs from the previous layer, and it delivers 24 feature maps to fully connected layers. Finally, we obtain thermal conductivity fields from CNN's outputs.

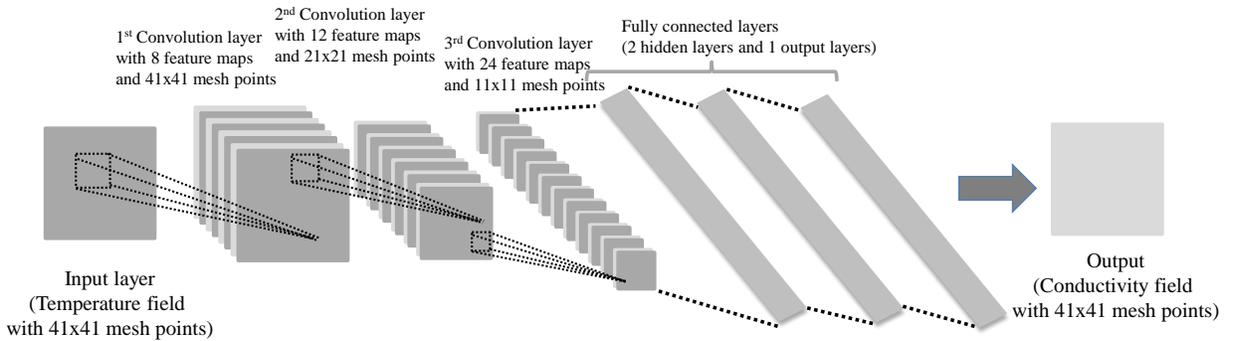

Fig. 16. Architecture of CNN-based thermal conductivity (adopted after LeCun) [48].

Learning is an optimization process, and we need to define a cost function based on distinct types of data to inform ML algorithms to tune NN hyperparameters. Eq. (11) defines the cost function where $N$, $y_{i,data}$, and $y_{i,model}$ are the total number of training data, $i^{th}$ training data, and $i^{th}$ model solution. To prevent overfitting, we add a regularization term in Eq. (11) where $i$, and $N_L$ denote the $i^{th}$ layer and total layer number. $\lambda$ is the regularization strength, and $\mathbf{W}$ is the matrix of total weights in $i^{th}$ layer. We implement NN-based thermal conductivity using Tensorflow [50] which is the DL framework developed by Google. Weights and biases of NNs are tuned based on data using Adam [51] algorithm.

$$E = \frac{1}{2N}\left[\sum_{i=1}^{N}\left(y_{i,\text{model}} - y_{i,\text{data}}\right)^2 + \sum_{i=1}^{N_L}\lambda_i \left\|\mathbf{W}_i\right\|^2\right] \tag{11}$$

### 4.4. Result analysis

#### 4.4.1. Comparing results by Type I and Type II ML using SET data

We refer the notation NN-A as NNs trained by a dataset A where NNs can be either CNNs or FNNs. Fig. 17 depicts the averaged RMSE by comparing validating datasets to Type I and Type II ML results. When we used the low-quality dataset (S1) to train FNNs, both Type I and Type II ML provided poor predictions. However, Type II ML does not query values from FNN-based thermal conductivity. The error does not accumulate during each iteration while solving PDEs. When predictions are away from the training domain, Type II ML exhibits better performance than Type I ML. When we trained FNNs by high-quality dataset (S2), both frameworks improve errors in predictions, but Type I ML displays better predictability than Type II ML.

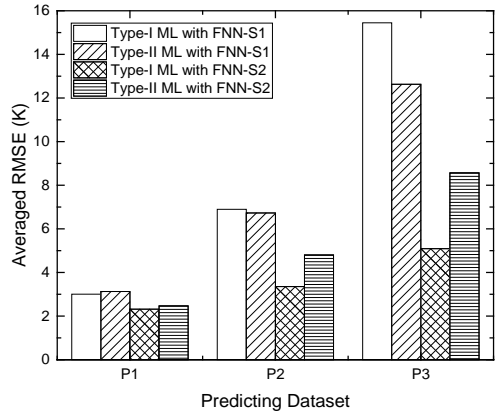

Fig. 17. Averaged RMSE by comparing the validating datasets, P1, P2, and P3, to Type I and Type II ML results using FNNs with training datasets, S1 and S2.

### 4.4.2. Comparing the results by Type III and Type V ML using IET data

Fig. 18 illustrates the averaged RMSE by comparing validating datasets to Type III and Type V ML results. Fig. 18(a) shows results by using Type III ML with FNN-based thermal conductivity. The results are improved as we increase the quantity of training data. Fig. 18(b) indicates that Type III ML with the CNN-T3 yields a lower error than the FNN-T3 result. However, CNN-based closures require more training data than FNN-based closures to exhibit good predictability. CNNs is efficient in the training. We performed simulations on NVIDIA TITAN Xp, and the training of Type III ML with the FNN-T3 took approximate 43.7 hours. On the contrary, Type III ML with the CNN-T3 only took about 1.2 hours to achieve a converged solution. Fig. 18(c) presents the averaged RMSE by Type V ML with FNNs. The RMSE cannot be improved while increasing the quantity of training data. The results imply that targets do not uniquely depend on inputs. Bishop [52] recognized this issue and solved it by mixture density networks.

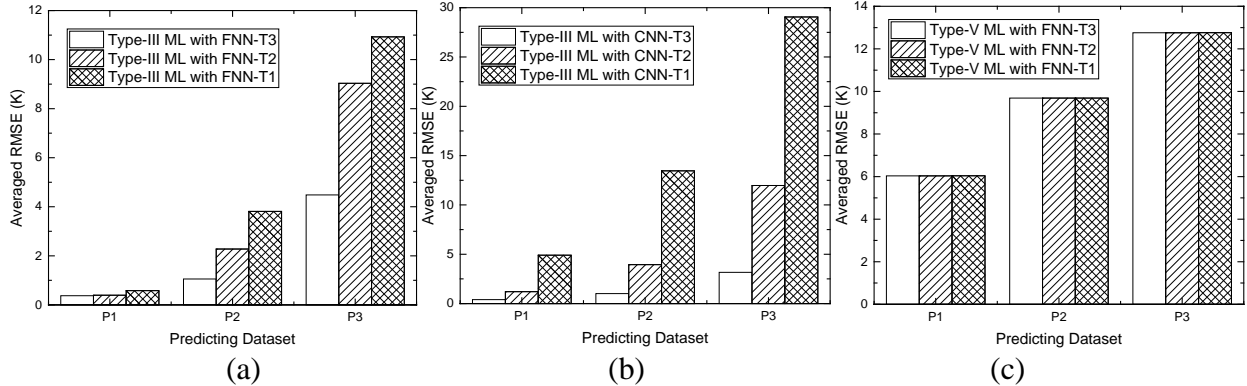

Fig. 18. Averaged RMSE by comparing the validating datasets, P1, P2, and P3, to the results obtained by (a) Type III ML using FNNs, (b) Type III ML using CNNs, and (c) Type V ML using FNNs with training datasets, T1, T2, and T3.

### 4.5. Lessons learned

When SET data are employed to Type I and Type II ML, the data quality strongly affects the accuracy of predictions. Type I ML requires to solve PDEs and query values from FNN-based closures for each iteration. If the data quality is low, errors accumulate in Type I ML, and Type II ML is more appropriate because Type II solves PDEs with fixed-field closures. When the data quality is high, Type I ML exhibits better performance than Type II ML. Type III ML trains a closure model that is embedded in PDEs by using IET data. The results are more accurate than Type I and Type II using SET data. Type III training requires more data than other two frameworks. When CNNs are used in Type III ML, the increase of training data significantly reduces the error in predictions. In the meanwhile, CNNs can accelerate the training in Type III ML by about 36 times faster than the training using FNNs. Type V ML results show an identifiability issue that indicates the selected input is not appropriate or different NNs should be used such as mixture density networks.

## 4.6. Summary

The case study indicates a preference for Type III ML. It can effectively utilize the field data, potentially generating more robust predictions than Type I, Type II, and Type V ML. Table 5 summarizes the properties of each ML framework based on the lessons learned in this study. The data quality needs to be high when NN-based closures are iteratively queried by Eq.(1). When the framework uses training data by IETs, the data quantity should be high to achieve predictions. The physics is conserved when the solution is constrained by Eq.(1). In this case study, Type V ML may require more input features than other ML frameworks. The selected input feature for Type V ML is insufficient to make the output uniquely depends on it.

Table 5. Properties of each ML framework for the heat conduction demonstration.

|  | Type I ML | Type II ML | Type III ML | Type V ML |
|---|---|---|---|---|
| Training data type | SET | SET | IET | IET |
| Data quantity requirement | Low | Low | High | High |
| Data quality requirement | High | Low | High | Low |
| Are NN-based closures iteratively queried while solving Eq.(1)? | Yes | No | Yes | No |
| Are solutions constrained by Eq.(1)? | Yes | Yes | Yes | No |
| Note | Type I is preferable when SET data quality is high. | Type II is preferable when SET data quality is low. | CNN-based closures are preferable. | There is an identifiability issue for the training. |

.

## 5. Two-phase flow modeling case study: Type I ML

### 5.1. Problem formulation

We use Type I ML to demonstrate how to construct a closure relation to close two-phase mixture models (TMMs) [53]. TMMs are convenient to deal with the phase appearance and disappearance, and it can consistently increase the fidelity by increasing field equations. Fig. 19 [54] summarizes the family of TMMs and its applicable problems. For example, the three-equation TMM assumes that the velocity, temperature, and pressure are homogeneous. It is not valid when the system includes the subcooled liquid. We can add field equations to extend applicable domains of mixture models. However, we need more closure relations to close field equations.

Thermal fluid simulation involves flow regime transitions, and this requires distinct closure models in each flow regime. The transitions complicate the scaling analysis since each experiment is valid in a particular domain and includes distinct uncertainties. Scalabrin, Condosta & Marchi [55, 56] utilized ANNs to build heat transfer models applied to a range of flow regimes for boiling flows. In this demonstration, we explore the hypothesis if the data-driven approach by using DL can construct the slip closure that is valid to a range of flow regimes in a vertical boiling channel. We start the investigation by using the three-equation TMM to predict a boiling channel task.

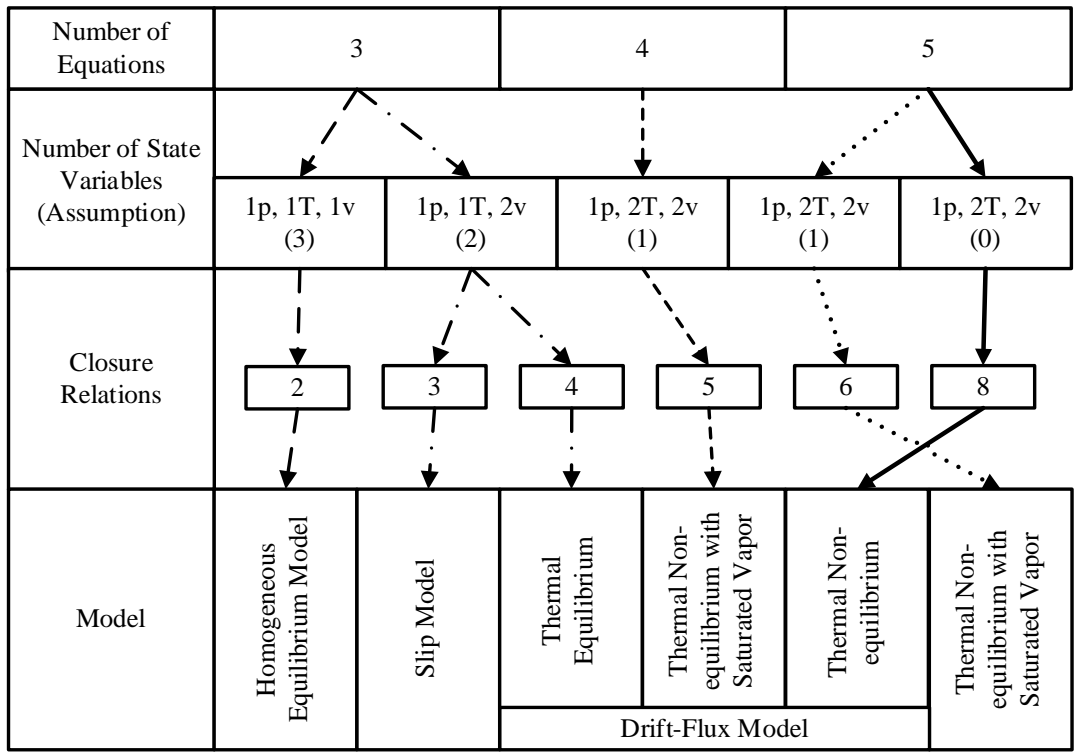

Fig. 19. The family of the TMM with various assumptions and closure relations (adopted after Wulff [54]).

## 5.2. Implementation

### 5.2.1. Implementation of the three-equation two-phase mixture model

We implement the three-equation TMM into the Modelica Standard Fluid Library [57]. Eq. (12)-(14) give the mass-energy-momentum conservation equation where $\rho$, $u$, $h$, and $v$, are the mixture density, internal energy, enthalpy, and velocity. The $l$, $v$, $A$, $P$, $\alpha$, $\tau_w$, and $P_w$ denote the liquid, vapor, area, pressure, void fraction, wall shear, and wetted perimeter. Eq. (15)-(16) are two-phase correction terms for the internal energy equation. Eq. (17) is the two-phase correction for the momentum equation. The mixture model requires a closure model for wall friction, but we do not implement two-phase correction terms for it due to the use of a fixed mass flow rate source. We further assume that there is no heat transfer, and the heat directly deposits into the fluid. As a result, we only need to develop the DL-based void-quality-slip closure to close the TMM, and we refer this model as the TMM-DL.

$$A\frac{\partial \rho}{\partial t} + \frac{\partial \rho v A}{\partial z} = 0 \tag{12}$$

$$A\frac{\partial \rho u}{\partial t} + \frac{\partial \rho h v A}{\partial z} = -P\frac{\partial v A}{\partial z} + q'(z) + E_{1,2\Phi} + E_{2,2\Phi} \tag{13}$$

$$\rho\frac{\partial v}{\partial t} + \rho v\frac{\partial v}{\partial z} = -\frac{\partial P}{\partial z} + \frac{\tau_w P_w}{A} - \rho g + M_{2\Phi} \tag{14}$$

$$E_{1,2\Phi} = -\frac{\partial}{\partial z}\left\{\frac{\alpha_l \alpha_v \rho_l \rho_v}{\rho}\left(u_v - u_l\right)\left(v_v - v_l\right)\right\}A \tag{15}$$

$$E_{2,2\Phi} = -P\frac{\partial}{\partial z}\left\{\frac{\alpha_l \alpha_v \rho_l \rho_v}{\rho}\left(\frac{1}{\rho_v} - \frac{1}{\rho_l}\right)\left(v_v - v_l\right)\right\}A \tag{16}$$

$$M_{2\Phi} = -\frac{\partial}{\partial z}\left\{\frac{\alpha_l \alpha_v \rho_l \rho_v}{\rho}\left(v_v - v_l\right)^2\right\}A \tag{17}$$

### 5.2.2. Implementation of deep NN-based slip closure

Eq. (18) shows the void-quality-slip closure [1] where $x$ is the fluid quality. The TMM requires a slip model that is the ratio of $v_g$ to $v_l$ to solve void fraction distributions. We construct DL-based slip models using deep FNNs (DFNNs) with a four-layer structure. Eq. (19) gives the formulation of slip models with three input parameters: the local pressure, local two-phase Reynolds number, and vapor Reynolds number. Eq. (20) defines the local two-phase Reynolds number [58] where $G_{2\Phi}$ is the two-phase mass flux and $\mu$ is the dynamic viscosity. Eq. (21) defines the vapor Reynolds number. FDNNs use sigmoidal activation functions with five HUs in each HL, and the model is implemented by Tensroflow [50] using the Adam [51] optimizer.

$$\alpha = \left(1 + \frac{1-x}{x}\frac{\rho_v}{\rho_l}S\right)^{-1} \tag{18}$$

$$S = DFNN(\mathbf{x}), with \quad \mathbf{x} = \left(P_{local}, \mathrm{Re}_{2\Phi,local}, \mathrm{Re}_{v,local}\right) \tag{19}$$

$$\mathrm{Re}_{2\Phi,local} = \frac{G_{2\Phi}D_{hyd}\left[x^2 + (1-x)^2(\rho_v/\rho_l)\right]}{\mu_v + \mu_l(1-x)(\rho_v/\rho_l)} \tag{20}$$

$$\mathrm{Re}_{v,local} = \frac{\rho_v v_v D_{hyd}}{\mu_v} \tag{21}$$

### 5.2.3. Implementation of two-phase flow modeling by Type I ML

Algorithm 5 shows the procedure of utilizing Type I ML to develop slip closures. Training data are obtained from the two-fluid model (TFM) [1]. We assume that there are invisibly tiny bubbles moving with the liquid when single-phase flows present. Therefore, we need to constrain the DL-based slip to have the minimum output equal to one. Then we implement the DL-based slip closure into the TMM for predictions. Eq. (22) defines the relative error to evaluate the performance of Type I ML.

---

**Algorithm 5 Type I ML for the system-level thermal fluid simulation.**

**Input:** training inputs ($P$, $\mathrm{Re}_{2\Phi}$, and $\mathrm{Re}_v$) by the TFM (*element 3* in Fig. 9)

   training target ($S = v_g/v_l$) (*element 4* in Fig. 9)

**Output:** TMM with DL-based slip closure for predictions (*element 7* in Fig. 9)

1: **for** i < maximum_iteration **do** (*element 5* in Fig. 9)

2:   // Build a slip model using DFNNs

3:   $S(P, \mathrm{Re}_{2\Phi}, \mathrm{Re}_v) \leftarrow DFNN$

4:   **for** all inputs $\in$ training datasets **do**

5:     Update parameters for each layer in DFNNs

6: Constrain the value of *DFNN* based on the model assumption: (*element 6* in Fig. 9)

   $DFNN \leftarrow \max(1, DFNN)$

7: Solve TMM with the DL-based slip closure: (*element 7* in Fig. 9)

   $\alpha = \left[1 + \frac{1-x}{x}\frac{\rho_v}{\rho_l}S(R, \mathrm{Re}_{2\Phi}, \mathrm{Re}_v)\right]^{-1}$

---

$$\varepsilon_r = \frac{\alpha_{TMM} - \alpha_{TFM}}{\alpha_{TFM}} \tag{23}$$

### 5.3. Manufacturing synthetic data for Type I ML

We use the TFM implemented by USNRC TRACE [59] to provide the synthetic data for building the void-quality-slip closure. We generate one baseline training dataset using the boiling channel characteristics given in Table 6. Then we create another eight validating datasets with distinct system characteristics.

Table 6. Characteristics of the boiling channel.

| Parameters | Values |
|---|---|
| Channel heat flux (J/sec-m$^2$) | 4.5436 x 10$^5$ |
| Outlet pressure (bar) | 67.73 |
| Coolant mass flux (kg/sec-m$^2$) | 1925.85 |
| Inlet temperature (K) | 550.93 |
| Heated diameter (m) | 0.0125 |
| Heated length (m) | 3.7084 |
| Flow area (m$^2$) | 1.41096 x 10$^{-4}$ |

### 5.4. Result analysis by using TMM-DL to predict various system characteristics

Fig. 20 depicts the void fraction comparison between the TMM-DL and TFM at the outlet for various system characteristics. When the system reaches the steady state at 720 seconds, a heater maintains a constant heat flux. Table 7 gives the relative errors of void fractions between the TFM and TMM-DL at the steady state. For the baseline, the relative error between the TMM-DL and TFM is 6.8% in the steady state. As we increase or reduce the power, the errors are below 5%. When we increase the mass flow rate by 20%, the error becomes 8.5%. However, the error goes down to 4.7% as we decrease the mass flow rate to 80%. When we increase the hydraulic diameter by the factor of 2, the error is merely 0.7%. The error becomes 3.2% as the hydraulic diameter decreases. For the 110% pressure case, the error is 4.2%, but a significant error (11.5%) occurs for the 95% pressure case.

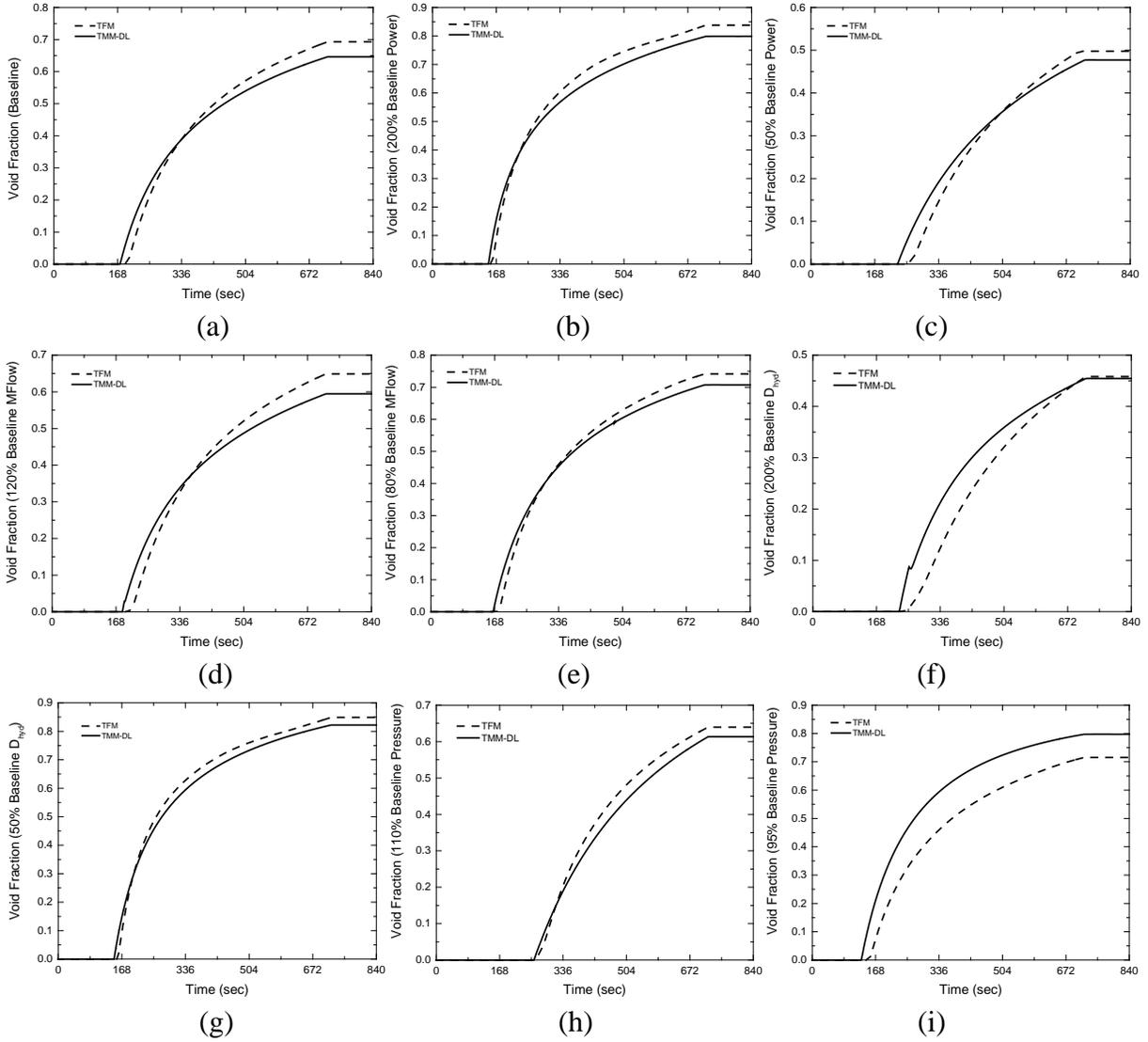

Fig. 20. Comparison of the void fraction at the pipe outlet between the TFM and TMM-DL for various system characteristics such as (a) the baseline, (b) 200% baseline power, (c) 50% baseline power, (d) 120% baseline mass flow rate (MFlow), (e) 80% baseline MFlow, (f) 200% baseline $D_{hyd}$, (g) 50% baseline $D_{hyd}$, (h) 110% baseline pressue, and (i) 95% baseline pressure.

Table 7. Relative errors of void fractions for the TMM-DL results comparing to the TFM results at the outlet under the steady-state condition.

| System Characteristics | Relative Errors of Void Fraction |
|---|---|
| Baseline | -6.8% |
| 200% Baseline Power | -4.7% |
| 50% Baseline Power | -4.2% |
| 120% Baseline MFlow | -8.5% |
| 80% Baseline MFlow | -4.7% |
| 200% baseline $D_{hyd}$ | -0.7% |
| 50% baseline $D_{hyd}$ | -3.2% |
| 110% Baseline Pressure | -4.2% |
| 95% Baseline Pressure | 11.5% |

### 5.5. Lessons learned

Type I ML can achieve the cost-effective development of the DL-based slip closure that is ubiquitous across flow regimes from the single-phase flow to flow boiling. The information of flow regime transitions implicitly inherits from data. For this case study, the DL-based slip closure exhibits the predictability beyond the training domain. Specifically, the mixture model can make predictions within a reasonable uncertainty range by comparing the results to the TFM for various system characteristics outside the training domain. However, further investigations are needed for a broader range of system conditions as well as for datasets generated by experiments. The presently synthetic data are limited to the TRACE simulation that is based on certain assumptions and models. It is noted that caution must be exercised in applying the synthetic data because they may have been biased by the previous calibration of models. The analysis conducted for the example of the TMM-DL modeling suggests that the performance of data-driven models may be affected by model biases. Therefore, the evaluation may be hampered by various sources of uncertainties including model forms and numerical errors. For instance, the drag force model in the TRACE TFM is inherited from the drift-flux model [59], whereas the tested mixture model is based on the phasic velocity slip formulation.

## 6. Concluding remarks

The present study is motivated by the growing interest and development of machine learning models in thermal fluid simulation. The trend is powered by the advent of data-intensive research methods, such as modern thermo-fluid experiments and high-fidelity numerical simulations, affordable computing (data processing) power and memory, and progress in machine learning methods, particularly in deep learning using multilayer neural networks. We introduced a classification of machine learning frameworks for thermal fluid simulation, including five types. The selection of the optimal ML framework is problem-dependent, and to a substantial extent, depends on characteristics of supporting data, including data source, data type, data quantity, and quality. Although examples of Type I and Type II models existed in the literature, their developments are still in infancy. While technically straightforward, both Type I and Type II models are limited to systems whose governing physics are amenable to "divide-to-conquer".

To the best knowledge of the authors, Type III (Physics-Integrated Machine Learning) framework was formulated and introduced for the first time in this study. In Type III, PDEs are involved in the training of machine-learning models, thus alleviating the requirements on the scale separation assumption, and potentially reducing the necessity on the physics decomposition. Correspondingly, Type III models present more stringent requirements on modeling and substantially higher computing resources for training. Based on insights from the case study performed, Type III ML has the highest potential in extracting the value from "big data" in thermal fluid research, while ensuring data-model consistency. There are technical challenges that need to be addressed before Type III models deliver their promises in practical thermal fluid simulation, namely, (i) complex interactions of ML-based closures with a system of PDEs (including discontinuity in hyperbolic systems); (ii) effect of the non-local character of ML-based models on PDE solution methods; and (iii) implementation and effect of multiple closure models, particularly in multiphase and thermal flows.

**Acknowledgments**

The support by the US Department of Energy via the Consortium for Advanced Simulation of Light Water Reactors (CASL), and NEUP Integrated Research Project is gratefully acknowledged. The authors thank NVIDIA Corporation for the Titan Xp GPU used for this research.

## Appendix A: Turbulent flow modeling examples: Type I and Type II ML

### A.1. Problem formulation

Reynolds-averaged Navier-Stokes (RANS) equations obtained by temporal averaging of Navier-Stokes equations require Reynolds stress to close the model. The linear eddy viscosity model (LEVM) has been widely used to represent Reynolds stress that leads to various mechanistic turbulence models [22] such as the Spalart-Allmaras, $k$-$\varepsilon$, and $k$-$\omega$ models. The models have been extensively studied, evaluated and calibrated for different flow characteristics with different degrees of accuracy. Consequently, performance of different models is limited in their calibration domain and exhibit high uncertainty in prediction regimes [35, 60].

With the advanced computing power, "first-principles" DNS and high-resolution LES have been used to generate high-fidelity turbulence data to inform turbulence modeling. Although not so named, Type I and Type II ML previously have been formulated and applied for data-driven turbulence modeling; e.g., in the work of Zhang & Duraisamy [20] and Ling, Kurzawski & Templeton [26]. Their implementation is analyzed with respect to the proposed frameworks in subsections 3.2 and 3.3, respectively.

### A.2. Implementation

#### A.2.1. Implementation of turbulent flow modeling by Type I ML

Zhang & Duraisamy [20] used the spatiotemporal function to modify the $k$-$\omega$ model that can inform RANS simulation by assimilating data from DNS. Fig. 21 depicts the application of Type I ML framework for data-driven turbulence modeling as proposed by Zhang & Duraisamy [20]. In correspondence with the structure described in Fig. 9, the procedure includes the following elements:

> *Element 1.* Assume scale separation is achievable such that DNS data ($\boldsymbol{\Psi_{DNS}}$) can be used to obtain a spatiotemporal function ($\boldsymbol{\alpha}$) in the $k$-$\omega$ model. Then collect RANS data ($\boldsymbol{\Psi_{RANS}}$) for computing candidates of flow features ($\boldsymbol{Q}$).

> *Element 2.* Average DNS data ($\boldsymbol{\Psi_{DNS}}$) to match the dimension of RANS data ($\boldsymbol{\Psi_{RANS}}$). Then scale the flow features ($\boldsymbol{Q}$) from element 1 as inputs for element 3

> *Element 3.* Select flow features ($\boldsymbol{Q}$) through the hill-climbing feature selection, and use the results as training inputs for element 5.

> *Element 4.* Compute the training targets, spatiotemporal factors ($\boldsymbol{\alpha}$), by solving the inverse problem using the turbulence kinetic energy equation and averaged $\boldsymbol{\Psi_{DNS}}$.

*Element 5.* Utilize an NN algorithm to capture the underlying correlation between flow features ($\boldsymbol{Q}$) and spatiotemporal factors ($\boldsymbol{\alpha}$). After the training, output the FNN-based spatiotemporal model, $FNN(\boldsymbol{Q}(\boldsymbol{\Psi_{RANS}}))$, to element 6.

*Element 6.* The $g(ML(\boldsymbol{Q}))$ is equal to $ML(\boldsymbol{Q})$ since there is no assumption made in the reference.

*Element 7.* Implement the FNN-based spatiotemporal model into the $k$-$\omega$ model, and solve RANS equations for predictions.

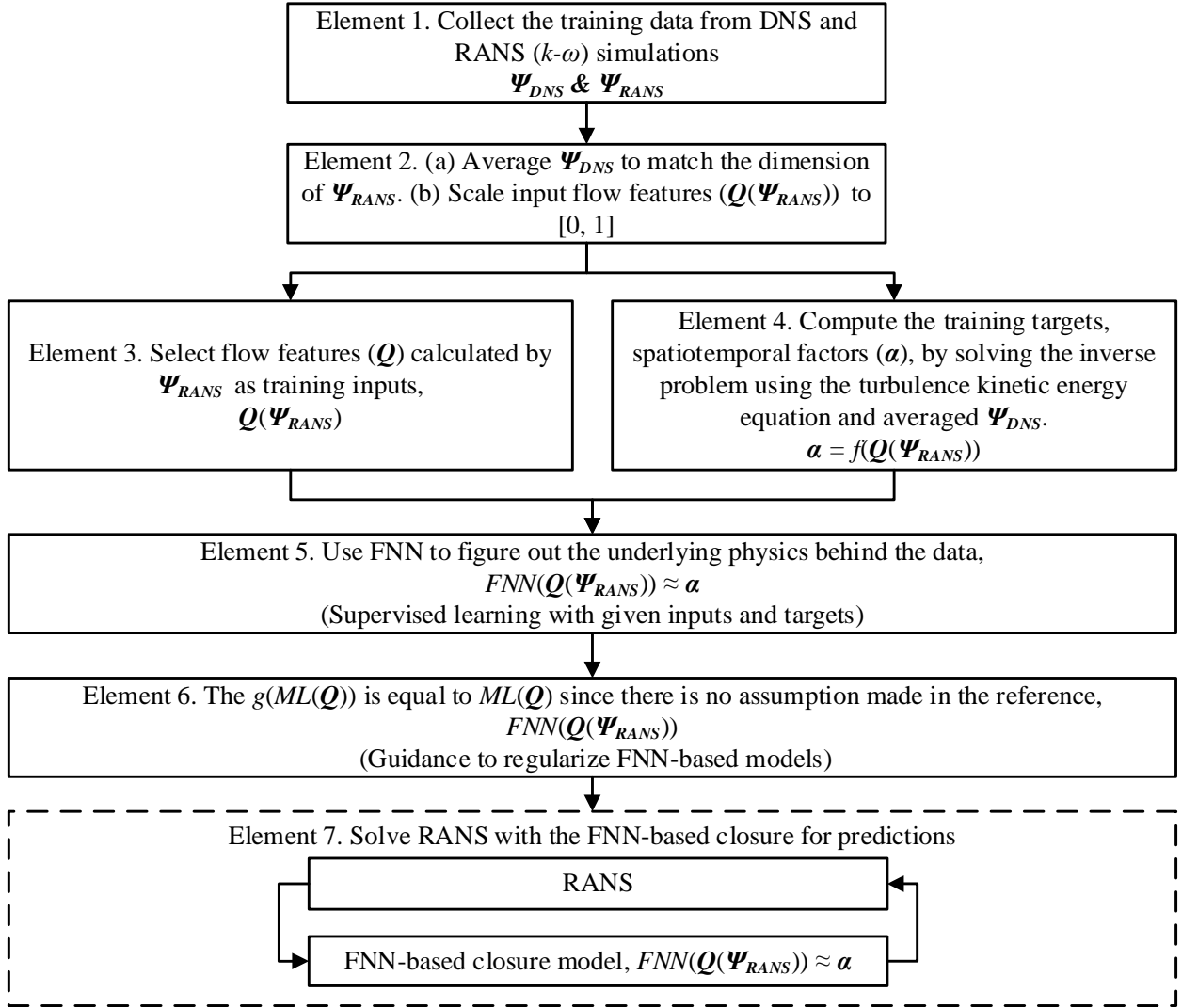

Fig. 21. Type I ML for data-driven turbulence modeling as proposed by Zhang & Duraisamy [20].

The study simulated 1D channel flow and used training datasets with friction Reynolds numbers ($Re_\tau$) [61] ranging from 180 to 4200. The result indicated that the reconstructed function ($\boldsymbol{\alpha}$) could be applied to the testing case with $Re_\tau$ equal to 2000.

### A.2.2. Implementation of turbulent flow modeling by Type II ML

Ling, Kurzawski & Templeton [26] utilized the ML-based Reynolds stress anisotropy tensors by tensor basis neural networks (TBNNs) to close RANS equations. Fig. 22 depicts the application of Type II ML framework for data-driven turbulence modeling as proposed by Ling, Kurzawski & Templeton [26]. In correspondence with the structure described in Fig. 10, the procedure includes the following elements:

*Element 1*. Perform RANS simulations with the *k-ε* model. The results ($\boldsymbol{\Psi_{RANS}}$) are prior solutions for training.

*Element 2*. Perform DNS simulations with identical system characteristics in element 1. The results ($\boldsymbol{\Psi_{DNS}}$) are baseline solutions for training.

*Element 3*. Average $\boldsymbol{\Psi_{DNS}}$ to match the dimension of $\boldsymbol{\Psi_{RANS}}$.

*Element 4*. Select five tensor invariants ($\boldsymbol{\lambda} = [\lambda_1, ..., \lambda_5]$) as training inputs for the input layer and ten isotropic basis tensors ($\boldsymbol{T} = [T_1, ..., T_{10}]$) as inputs for the tensor input layer by $\boldsymbol{\Psi_{RANS}}$. $\boldsymbol{\lambda}(\boldsymbol{\Psi_{RANS}})$ and $\boldsymbol{T}(\boldsymbol{\Psi_{RANS}})$ are training inputs to element 6. It is noted that the $\boldsymbol{\lambda}$ and $\boldsymbol{T}$ can be computed from the non-dimensionalized strain rate ($\boldsymbol{S}$) and rotation rate tensors ($\boldsymbol{R}$).

*Element 5*. Compute Reynolds stress anisotropy tensors ($\boldsymbol{b}$) by averaged $\boldsymbol{\Psi_{DNS}}$ as the training targets that can supervise NN algorithms to learn from data.

*Element 6* Use NN algorithms to represent the underlying correlation of the non-dimensionalized strain rate ($\boldsymbol{S}$), rotation rate tensors ($\boldsymbol{R}$) and Reynolds stress anisotropy tensors ($\boldsymbol{b}$). After the training, output the TBNN-based Reynolds stress anisotropy tensor, *TBNN*($f(\boldsymbol{\lambda}(\boldsymbol{\Psi_{RANS}}))$, $\boldsymbol{T}(\boldsymbol{\Psi_{RANS}})$), to element 8

*Element 7*. Execute a new RANS (*k-ε*) simulation ($\boldsymbol{\Psi'_{RANS}}$) with different system characteristics. Then use the solution to obtain $\boldsymbol{\lambda}$ and $\boldsymbol{T}$ as inputs to element 8.

*Element 8*. Use $\boldsymbol{\lambda}$ and $\boldsymbol{T}$ from element 7 as inputs to query values from the TBNN-based Reynolds stress anisotropy tensor model, *TBNN*($f(\boldsymbol{\Psi_{RANS}})$, $\boldsymbol{T}(\boldsymbol{\Psi_{RANS}})$). Output the Reynolds stress anisotropy tensor as fixed fields to element 9.

*Element 9*. Implement the results from element 8 into the RANS solver, SIERRA Fuego [62], for predictions.

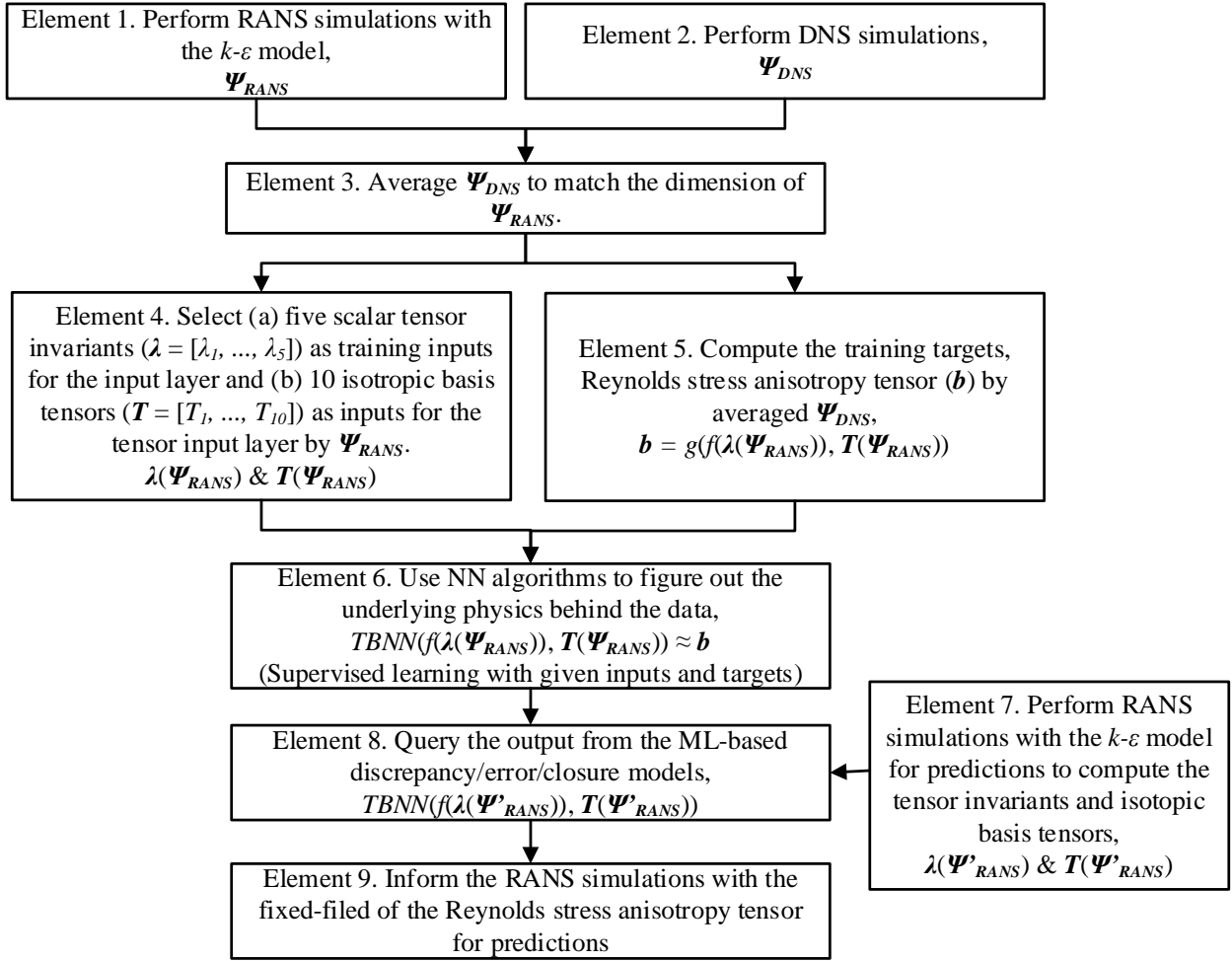

Fig. 22. Type II ML for data-driven turbulence modeling as proposed by Ling, Kurzawski & Templeton [26].

Two flow cases had been tested including turbulent duct flow ($Re_b = 2000$) and flow over a wavy wall ($Re = 6850$). The TBNN was trained by six cases with various Reynolds number given in Table 8. The results indicated that the TBNN with embedded Galilean invariance could be used for Reynolds stress anisotropy predictions which is better than generic NNs. Notably, the TBNN yields more accurate predictions than the LEVM.

Table 8. High-fidelity simulations for training the TBNN

| Duct flow [63] | Channel flow [64] | Inclined jet in cross-flow [65] | Perpendicular jet in cross-flow [66] | Flow around a square cylinder [67] | Flow through a converging-diverging channel flow [68] |
|---|---|---|---|---|---|
| $Re_b = 3500$ | $Re_\tau = 590$ | $Re_{jet} = 5000$ | $Re_{jet} = 5000$ | $Re = 21400$ | $Re_\tau = 600$ |

Kutz [69] suggested that DNNs have potential to bring a paradigm shift in modeling of complex flows thanking their capability to capture multiscale features from data. He indicated that ROMs for fluids based on the singular value decomposition have difficulties to capture transient and multiscale phenomena as well as invariances due to scaling. On the contrary, DNNs can capture multi-scale features [70] through its hierarchy. Although DNNs can predict trends in data well, it is a challenge for DNNs to generate readily interpretable physical models.

*A.3. Lessons learned*

Zhang & Duraisamy's [20] work belongs to Type I ML. They showed that the spatiotemporal function allowed the $k$-$\omega$ model to assimilate data. However, the method requires Boussinesq hypothesis that limits the function form to eddy viscosity models. It requires extensive demonstration to show the applicability of the method regarding flows in a different geometry and regime.

Ling, Kurzawski & Templeton's [26] work belongs to Type II ML. They demonstrated the TBNN captured the invariant of Reynolds stress modeling for various flows. The work used fixed fields of the DL-based Reynolds stress to close RANS equations. The authors also mentioned that the stress model should be iteratively queried while solving RANS equations. The demonstrations use steady-state cases to show that the TBNN can improve RANS predictions in different geometries and at distinct Reynolds numbers. Type II ML exists an open question about what the magnitude of errors can be before it is too late to bring a prior solution to a baseline.